\documentclass[review,5p,times,twocolumn]{elsarticle}

\usepackage{ifthen}
\usepackage{xcolor}
\usepackage{float}
\usepackage{array}
\usepackage{adjustbox}
\usepackage{caption}
\usepackage{subcaption}
\usepackage{rotating}
\usepackage{xspace}
\usepackage[normalem]{ulem}
\usepackage{amssymb}
\usepackage{amsmath}
\usepackage{comment}    
\usepackage{tabularx}
\usepackage{csvsimple,supertabular}
\usepackage{graphicx}
\usepackage{booktabs}

\usepackage{enumitem} 

\newcommand{\id}[1]{$-$Id: scgPaper.tex 32478 2010-04-29 09:11:32Z oscar $-$}

\newboolean{showedits}
\setboolean{showedits}{false} 
\ifthenelse{\boolean{showedits}}
{
	\newcommand{\del}[1]{\textcolor{red}{\sout{#1}}} 
	\newcommand{\nbe}[3]{
		{\colorbox{#3}{\bfseries\sffamily\scriptsize\textcolor{white}{#1}}}
		{\textcolor{#3}{\sf\small$\blacktriangleright$\textit{#2}$\blacktriangleleft$}}}
}{
	\newcommand{\del}[1]{} 
	
	\newcommand{\nbe}[3]{}
}

\newboolean{showcomments}
\setboolean{showcomments}{false} 
\ifthenelse{\boolean{showcomments}}
 {
 	\newcommand{\nbc}[3]{
 		{\colorbox{#3}{\bfseries\sffamily\scriptsize\textcolor{white}{#1}}}
		{\textcolor{#3}{\sf\small$\blacktriangleright$\textit{#2}$\blacktriangleleft$}}}
	
 }{
 	\newcommand{\nbc}[3]{}
 	
 }
\usepackage[most]{tcolorbox}
\ifthenelse{\boolean{showedits}}
{
  \newtcolorbox{inserted}{%
       title=Inserted text:,
       colframe=blue,colback=blue!5!white,
       breakable,
       leftrule=0mm, 
       bottomrule=0mm,
       rightrule=0mm,
       toprule=0mm,
       arc=0mm, outer arc=0mm,
       oversize
  }
  \newtcolorbox{deleted}{%
       title=Deleted text:,
       colframe=red,colback=red!5!white,
       breakable,
       leftrule=0mm, 
       bottomrule=0mm,
       rightrule=0mm,
       toprule=0mm,
       arc=0mm, outer arc=0mm,
       oversize
  }
  \newtcolorbox{refactored}{%
       title=Rewritten text:,
       colframe=blue,colback=red!5!white,
       breakable,
       leftrule=0mm, 
       bottomrule=0mm,
       rightrule=0mm,
       toprule=0mm,
       arc=0mm, outer arc=0mm,
       oversize
  }
}{

}
\newcommand{\seclabel}[1]{\label{sec:#1}}
\newcommand{\secref}[1]{\autoref{sec:#1}}
\newcommand{\figlabel}[1]{\label{fig:#1}}
\newcommand{\figref}[1]{\autoref{fig:#1}}
\newcommand{\tablabel}[1]{\label{tab:#1}}
\newcommand{\tabref}[1]{\autoref{tab:#1}}

\newcommand{\commented}[1]{}

\newcommand{\eg}{\emph{e.g.,}\xspace}
\newcommand{\ie}{\emph{i.e.,}\xspace}
\newcommand{\etal}{\emph{et al.}\xspace}
\newcommand{\etc}{\emph{etc.}\xspace}
\renewcommand{\rq}[1]{\textbf{RQ\textsubscript{#1}}} 
\newcommand{\qas}{{QAs}\xspace}
\newcommand{\qa}{{QA}\xspace}

\usepackage{tikz}
\tikzstyle{mybox} = [draw=black, very thick, rectangle, rounded corners, inner ysep=5pt, inner xsep=5pt]
\newcommand{\ballnumber}[1]{\tikz[baseline=(myanchor.base)] \node[circle,fill=.,inner sep=1pt] (myanchor) {\color{-.}\bfseries\footnotesize \textsf{#1}};}
\usepackage[T1]{fontenc}
\usepackage{textcomp}
\usepackage{listings}
\definecolor{source}{gray}{0.9}
\newcommand{\rqI}{What types of comments do researchers focus on when assessing comment quality?}
\newcommand{\rqII}{What \qas do researchers consider in assessing comment quality?}
\newcommand{\rqIII}{Which tools and techniques do researchers use to assess comment \qas?}
\newcommand{\rqIV}{What kinds of contribution do studies often make?}
\newcommand{\rqV}{How do researchers evaluate their comment quality assessment studies?}

\newcommand{\finding}[2]{\vspace{2mm}
\noindent\fbox{\parbox{0.98\linewidth}{\fontsize{9}{10}\selectfont
\textbf{Finding #1.} #2 }}\\}

\usepackage[pdftex,colorlinks=true,pdfstartview=FitV,linkcolor=black,citecolor=black,urlcolor=black,bookmarks=false]{hyperref}

\usepackage{fp}
\usepackage{numprint}
  \npthousandsep{,}
  \npdecimalsign{.}
\newcommand{\FPuse}[1]{\FPeval{\result}{#1}{\result}}

\newcommand{\asText}[1]{\textcolor{blue}{#1}\xspace}

\newcommand{\numQA}{21}
\newcommand{\additionalQA}{10}

\newcommand{\yearBegin}{2011}
\newcommand{\yearEnd}{2020}

\newcommand{\numConfsSearchAll}{110}
\newcommand{\numJournalsSearchAll}{85}
\newcommand{\numConfsSearchSelected}{20}
\newcommand{\numJournalsSearchSelected}{6}

\newcommand{\numPapersFomCrawl}{17554}

\newcommand{\numPapersReviewed}{2353}
\newcommand{\numPapersReviewedWithoutSnowball}{2043}

\newcommand{\numPapersCandidateWithoutSnowball}{71}

\newcommand{\numPapersRelevantWithoutSnowball}{30}
\newcommand{\numPapersRelevant}{47}

\newcommand{\numSnowballCitationsTotal}{2610}
\newcommand{\numSnowballCitationsUnique}{1624}
\newcommand{\numSnowballCitationsSelected}{741}
\newcommand{\numSnowballRefsTotal}{3369}
\newcommand{\numSnowballRefsUnique}{2080}
\newcommand{\numSnowballRefsSelected}{2021}

\newcommand{\numPapersSnowballAddedAll}{311}
\newcommand{\numPapersSnowballSelected}{39}
\newcommand{\numPapersSnowballRelevant}{17}

\journal{Journal of Systems and Software}

\begin{document}
\begin{frontmatter}

\title{A Decade of Code Comment Quality Assessment: A Systematic Literature Review}

\author[SCG]{Pooja Rani}
\ead{pooja.rani@unibe.ch}
\author[USI]{Arianna Blasi}
\ead{arianna.blasi@usi.ch}
\author[SCG]{Nataliia Stulova}
\ead{nataliia.stulova@inf.unibe.ch}
\author[Zahw]{Sebastiano Panichella}
\ead{panc@zhaw.ch}
\author[IMDEA]{Alessandra Gorla}
\ead{alessandra.gorla@imdea.org}
\author[SCG]{Oscar Nierstrasz}
\ead{oscar.nierstrasz@unibe.ch}

\address[SCG]{Software Composition Group, University of Bern, Bern, Switzerland}
\address[USI]{Università della Svizzera italiana, Lugano, Switzerland}
\address[Zahw]{Zurich University of Applied Science, Zurich, Switzerland}
\address[IMDEA]{IMDEA Software Institute, Madrid, Spain}

\begin{abstract}

  Code comments are important artifacts in software systems and play a
  paramount role in many software engineering (SE) tasks related to
  maintenance and program comprehension.  However, while it is widely
  accepted that high quality matters in code comments just as it
  matters in source code, \emph{assessing} comment quality in practice
  is still an open problem. First and foremost, there is no unique
  definition of quality when it comes to evaluating code comments. The
  few existing studies on this topic rather focus on specific
  attributes of quality that can be easily quantified and
  measured. Existing techniques and corresponding tools may also focus
  on comments bound to a specific programming language, and may only
  deal with comments with specific scopes and clear goals (\eg Javadoc
  comments at the method level, or in-body comments describing TODOs to be
  addressed).

  In this paper, we present a Systematic Literature Review (SLR) of the last decade of research in SE to answer the following research questions: (i) What \emph{types of comments} do researchers focus on when assessing comment quality? (ii) What \emph{quality attributes}
(\qas) do they consider? (iii) Which \emph{tools and techniques} do they use to assess comment quality?, and (iv) How do they \emph{evaluate} their studies on comment quality assessment in general?
  
  Our evaluation, based on the analysis of \asText{\numPapersReviewed} papers and the actual review of \asText{\numPapersRelevant} relevant ones,
  shows that (i) most studies and techniques focus on comments in Java code, thus may not be generalizable to other languages, and (ii) the analyzed studies focus on four main \qas of a total of \numQA~\qas identified in the literature, with a clear predominance of checking \emph{consistency} between comments and the code. 
  We observe that researchers rely on manual assessment and specific heuristics rather than the automated assessment of the comment quality attributes, with evaluations often involving surveys of students and the authors of the original studies but rarely professional developers.

\end{abstract}

\begin{keyword}
  {code comments\sep documentation quality\sep systematic literature review}
\end{keyword}

\end{frontmatter}


\section{Introduction}
\seclabel{introduction}
Software systems are often written in several programming languages~\cite{Abid20a}, and interact with many hardware devices and software components~\cite{Lehm97a,Torn18a}.
To deal with such complexity and to ease maintenance tasks, developers tend to document their software with various artifacts, such as design documents and code comments~\cite{Souz05a}. Several studies have demonstrated that \emph{high quality} code comments can support developers in software comprehension, bug detection, and program maintenance activities~\cite{Deke09a,McMi10a,Tan07c}.  
However, code comments are typically written using natural language sentences, and their syntax is neither imposed by a programming language's grammar nor checked by its compiler.
Additionally, static analysis tools and linters provide limited syntactic support to check comment quality.
Therefore, writing high-quality comments and maintaining them in projects is a responsibility mostly left to developers~\cite{Alla14a,Kern99a}.

The problem of \emph{assessing the quality of code comments} has gained a lot of attention from researchers during the last decade~\cite{Kham10a,Stei13b,Rato17a,Pasc17a,Wen19a}.
Despite the research community's interest in this topic, there is no clear agreement on what quality means when referring to code comments.
Having a general definition of quality when referring to code comments is hard, as comments are diverse in purpose and scope.

\textbf{Problem Statement}. Maintaining high-quality code comments is vital for software evolution activities, however, \textit{assessing the overall quality of comments is not a trivial problem}. 
As developers use various programming languages, adopt project-specific conventions to write comments, embed different kinds of information in a semi-structured or unstructured form~\cite{Padi09a,Pasc17a}, and lack quality assessment tools for comments, ensuring comment quality in practice is a complex task.
Even though specific comments follow all language-specific guidelines in terms of syntax, it is still challenging to determine automatically whether they satisfy other quality aspects, such as whether they are consistent or complete with respect to the code or not~\cite{Zhou17a}.
There are various such aspects, \eg readability, content relevance, and correctness that should be considered when assessing comments, but tools do not support all of them.
Therefore, a comprehensive study of the specific attributes that influence code comment quality and techniques proposed to assess them is essential for further improving comment quality tools.

Previous mapping and literature review studies have collected numerous quality attributes (\qas) that are used to assess the quality of software documentation based on their importance and effect on the documentation quality.
Ding~\etal~\cite{Ding14a} focused specifically on software architecture and requirement documents,
while Zhi~\etal~\cite{Zhi15a} analyzed code comments along with other types of documentation, such as requirement and design documents.
They identified 16 \qas that influence the quality of software documentation.
However, the identified \qas are extracted from a body of literature concerning relatively old studies (\ie studies conducted prior to the year \asText\yearBegin) and are limited in the context of code comments.
For instance, only 10\% of the studies considered by Zhi~\etal concern code comments.
Given the increasing attention that researchers pay to comment quality assessment, it is essential to know which \qas, tools and techniques they propose to assess code comment quality.

To achieve this objective, we perform an SLR on studies published in the last decade, \ie \asText{\yearBegin}-\asText{\yearEnd}.
We review \asText{\numPapersReviewed} studies and find \asText{\numPapersRelevant} to be relevant to assessing comment quality.
From these we extract the programming language, the types of analyzed comments, \qas for comments, techniques to measure them, and the preferred evaluation type to validate their results.

We observe that
(i) most of the studies and techniques focus on comments in Java code,
(ii) many techniques that are used to assess \qas are based on heuristics and thus may not be generalizable to other languages,
(iii) a total of \asText{\numQA} \qas are used across studies, with a clear dominance of \emph{consistency}, \emph{completeness}, \emph{accuracy}, and \emph{readability}, and
(iv) several \qas are often assessed manually rather than with the automated approaches.
We find that the studies are typically evaluated by measuring performance metrics and surveying students rather than by performing validations with practitioners.
This shows that there is much room for improvement in the state of the art of comment quality assessment.

The \textbf{contributions} of this paper are:
\begin{enumerate}[label=\roman*)]
  \item an SLR of a total of \asText{\numPapersReviewed} papers, of which we review the \asText{\numPapersRelevant} most relevant ones, focusing on \qas mentioned and research solutions proposed to assess code comment quality,
  \item  a catalog of \asText\numQA~\qas of which four \qas are often investigated, while the majority is rarely considered in the studies, and of which \asText\additionalQA are new with respect to the previous study by Zhi \etal~\cite{Zhi15a},
  \item a catalog of methods used to measure these \asText\numQA~\qas in research studies, 
  \item  an overview of the approaches and tools proposed to assess comment quality, taking into account the types of comments and the programming languages they consider,
  \item a discussion of the challenges and limitations of approaches and tools proposed to assess different and complementary comment \qas, and
  \item  a publicly available dataset including all validated data, and steps to reproduce the study in the replication package.\footnote{\url{https://doi.org/10.5281/zenodo.4729054}}
\end{enumerate}

\textbf{Paper structure}. The rest of the paper is organized as follows.
In \secref{study-design} we highlight our motivation and rationale behind each research question, and we present our methodology, including the different steps performed to answer our research questions.
In \secref{results} we report the study results.
We discuss the results in \secref{discussion} and their implications and future direction in \secref{implication-future-work}.
We highlight the possible threats to validity for our study in \secref{Threats-to-validity}.
Then \secref{Related-work} summarizes the related work, in relation to the formulated research questions. Finally, \secref{conclusion} concludes our study, outlining future directions.

\section{Study Design}
\seclabel{study-design}

The main objective of our study is to present an overview of the state
of the art in assessing the quality of code comments. 
Specifically, we aim to highlight the \qas mentioned in the literature, and the
techniques used so far to assess comment quality.
To this end, we carry out an SLR, following the widely accepted guidelines of Kitchenham
\etal~\cite{Kitc07a} and Keele~\cite{Keel07a}.  
The first step in this direction is to specify the research questions related to the topic of interest~\cite{Kitc07a}. 
The following steps focus on finding a set of relevant studies that are related to the research questions based on an unbiased search strategy.

\subsection{Research questions}
\label{subsec:research-questions}
Our \textit{goal} is to foster research that aims at 
building code comment assessment tools.
To achieve this goal, we conduct an SLR, investigating the literature of the last decade to identify comment related \qas and solutions that address related challenges. 
We formulate the following research questions:
\begin{itemize}
\item \rq{1}: \emph{\rqI}
 \emph{Motivation}: Comments are typically placed at the beginning of a file, usually to report
  licensing or author information, or placed preceding a class or function to document the overview of a class or function and its implementation details.
  Depending on the specific type of comment used in source code and the specific programming language, researchers may use different techniques to assess them. These techniques may not be generalizable to other languages.
  For example, studies analyzing class comments in object-oriented programming languages may need extra effort to generalize the comment assessment approach to functional programming languages.
  We, therefore, investigate the comment types researchers target.

  \item \rq{2}: \emph{\rqII} \\
  \emph{Motivation}: \qas may solely concern syntactic aspects of the comments (\eg syntax of comments), writing style (\eg grammar), or content aspects (\eg consistency with the code).
  Researchers may use different terminology for the same \qa and thus these terms must be mapped across studies to obtain a unifying view of them, for instance, if the \emph{accuracy} \qa is defined consistently across studies or another terminology is used for it.
  We collect all the possible \qas that researchers refer to and map them, if necessary, following the methodology of Zhi \etal.
  Future studies that aim to improve specific aspects of comment quality evaluation can use this information to design their tools and techniques.

  \item \rq{3}: \emph{\rqIII} \\
  \emph{Motivation}: Researchers may assess \qas manually, or may use
  sophisticated tools and techniques based on simple heuristics or complex machine learning (ML) to assess them automatically.
  We aim to identify if there are clear winning techniques for this domain and collect various metrics and tools used for this purpose.

  \item \rq{4}: \emph{\rqIV}\\
  \emph{Motivation}: 
Engineering researchers usually motivate their research based on the utility of their results.
  Auyang clarifies that engineering aims to apply scientific methods to real world problems~\cite{Auya06a}.  
  However, software engineering currently lacks validation~\cite{Zelk97a}.
  With this question, we want to understand what types of solution researchers contribute to improving automatic comment quality assessment, such as metrics, methods, or tools.
  This RQ can provide insight into specific kinds of solutions for future work.

  \item \rq{5}: \emph{\rqV}\\
  \emph{Motivation}: Researchers may evaluate their comment assessment approaches, \eg by surveying developers, or by using a dataset of case studies.
  However, how often they involve professional developers and industries in such studies is unknown.
\end{itemize}

\subsection{Search Strategy}
\label{subsec:search-strategy}
After formulating the research questions, the next steps focus on finding relevant studies that are related to the research questions. 
In these steps, we 
\begin{enumerate}
  \item construct search keywords in subsection \ref{subsec:search-keywords},
  \item choose the search timeline in subsection \ref{subsec:search-timeline},
  \item collect sources of information in subsection \ref{subsec:autom-data-coll},
  \item retrieve studies in subsection \ref{subsec:data-retrieval},
  \item select studies based on the inclusion/exclusion criteria in subsection \ref{subsec:data-selection}, and
  \item evaluate the relevant studies to answer the research questions in subsection \ref{subsec:data-evaluation}.
\end{enumerate} 

\begin{figure*}[ht!]
  \centering
  \includegraphics[width=\textwidth]{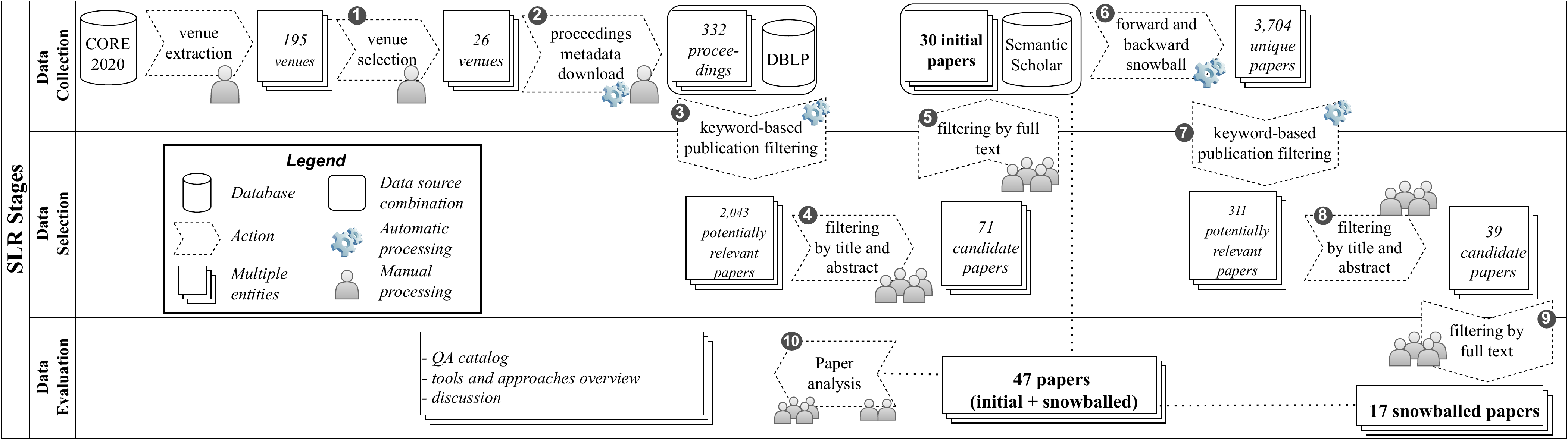}
  \caption{SLR stages to collect relevant papers}
  \figlabel{paper-selection-process}
\end{figure*}

\subsubsection{Search Keywords}
\label{subsec:search-keywords}
Kitchenham \etal recommended formulating individual facets or search units based on the research questions~\cite{Kitc07a}. These search units include abbreviations, synonyms and other spellings, and they are combined using boolean operators.
Pettricrew \etal suggested PIO (population, interventions, and outcome) criterion to define such search units~\cite{Pett08a}.

The \emph{populations} include terms related to the standards.
We first examine the definitions of \emph{documentation} and \emph{comment} in \emph{IEEE Standard Glossary of Software Engineering Terminology} (IEEE Standard 610.12-1990) to collect the main keywords. 
According to the definition, we identify the keywords \emph{comment}, \emph{documentation}, and \emph{specification} and add them to the set \emph{$K_{1}$}.
We further add frequently mentioned comment-related keywords, such as \emph{API}, \emph{annotation}, and \emph{summar} to the set \emph{$K_{1}$}.

The \emph{interventions} include terms that are related to software methodology, tools, technology, or procedures.
With respect to quality assessment, we define the intervention keywords to be \emph{quality}, \emph{assess}, \emph{metric}, \emph{measure},
\emph{score}, \emph{analy}, \emph{practice}, \emph{structur},
\emph{study}, or \emph{studied} and add them to the set \emph{$K_{2}$}.

Note that we add common variations of the words manually, for example, we add ``summar'' keyword to the set to cover both ``summary'' and ``summarization''.
We do not use any NLP libraries to stem words due to two main reasons, (i) to reduce the noisy matches, and (ii) the words from the title and abstract of the papers are not preprocessed (stemmed or lemmatized), therefore stemming the keywords might not find the exact or prefix matches.
For example, using the porter stemming approach, the word ``study'' will be stemmed to ``studi'' and we might miss the papers with ``study'' word. 
To avoid such cases, we add common variations of this word \emph{study} and \emph{studied} to our search keywords.

The \emph{outcomes} include terms that are related to factors of significance to developers (\eg reduced cost, reduced time to assess quality). Since it is not a required unit to restrict the search scope, and our focus is on all kinds of quality assessment approaches, we exclude the outcomes in our search keywords.
However, to narrow down our search and exclude irrelevant papers, such those about code reviews or testing, or non-technical papers, we formulate another set of keywords, \emph{$K_{3}$}.
In this set, we include \emph{code review}, \emph{test}, \emph{keynote}, \emph{invited}, and \emph{poster}, to exclude entries
of non-technical papers that were not filtered out using the
heuristics on the number of pages.

\begin{table}[t]
  \caption{keywords selected according to PIO criterion}
  \tablabel{PIO-keywords}
  \begin{tabular}{l|p{0.60\linewidth}}
    \hline
  \textbf{Criteria} &  \textbf{keywords}
  \\ \hline
  Populations ($K_{1}$)
  & \emph{comment}, \emph{documentation}, \emph{specification}, \emph{API}, \emph{annotation}, and \emph{summar}
  \\
  Interventions ($K_{2}$)
  & \emph{quality}, \emph{assess}, \emph{metric}, \emph{measure},
  \emph{score}, \emph{analy}, \emph{practice}, \emph{structur},
  \emph{study}, and \emph{studied}
  \\
  \end{tabular}
  \end{table}

Hence, using the final set of keywords (also given in \tabref{PIO-keywords}), we select a paper if its title and abstract match the keywords from $K_{1}$ and $K_{2}$ but not from $K_{3}$ where the prefix function is used to match the keywords in the paper. 

\subsubsection{Timeline}
\label{subsec:search-timeline}
We focus our SLR on the last decade (\ie January \asText\yearBegin-December \asText\yearEnd) since Zhi \etal investigated the works on software documentation quality --- including code comments --- from \asText{1971} to \asText\yearBegin~\cite{Zhi15a}. 
Our results can thus be used to observe the evolution of comment quality assessment, but, more importantly, they naturally complement the existing body of knowledge on the topic.

We then proceed to the main steps \ie retrieving the paper data, selecting venues, and identifying the relevant papers for our comment context.

\subsubsection{Data collection}
\label{subsec:autom-data-coll}
Concretely, our data collection
approach comprises three main steps, \ie literature data collection,
data selection, and data evaluation, which we sketch in~\figref{paper-selection-process} and present in further detail as follows:

We now describe how we automatically collect the data from the
literature, explaining the rationale behind our selection of venues
and our automatic keyword-based filtering to identify the likely
relevant papers regarding comment quality assessment.
We justify the need for another step of data gathering based on the snowball approach in Section~\ref{subsec:data-selection}.
Finally, we present our criteria for the careful evaluation of the
relevant papers in Sec \ref{subsec:data-evaluation}.

\textbf{Venue Selection.}
Code comment analysis, generation, usage, and maintenance are of primary interest to the SE research community.
Thus, in order to systematically review the literature on the comment quality assessment, we start by focusing on the SE venues.
We use the latest 2020 updated version of the conference and journal database of the CORE ranking portal as a primary data source to identify all the potentially relevant SE
venues.\footnote{\url{https://www.core.edu.au/conference-portal}}
The portal provides assessments of major conferences and journals in the computing disciplines, and it is a well-established and regularly-validated registry maintained by the academic community.
We extract all ranked journals in SE (search code 803) from the CORE portal\footnote{\url{http://portal.core.edu.au/jnl-ranks/?search=803&by=for&source=CORE2020&sort=arank&page=1} accessed on 25 Mar, 2021} and all top conferences and workshops in the SE field (search code 4612).\footnote{\url{http://portal.core.edu.au/conf-ranks/?search=4612&by=for&source=CORE2020&sort=arank&page=1} accessed on 25 Mar, 2021}
 This process gives us an initial list of \asText\numJournalsSearchAll journal and \asText\numConfsSearchAll conference venues.
We select in step\,\ballnumber{1}
\asText{\FPuse{round(\numConfsSearchSelected+\numJournalsSearchSelected, 0)}} software engineering (SE) conferences and journals from \asText{\FPuse{round(\numConfsSearchAll+\numJournalsSearchAll, 0)}} candidate venues based on the
likelihood of finding relevant papers in their proceedings.

We focus on A* and A conferences and journals, and add conferences of rank B or C if they are co-located with previously selected A* and A conferences to have venues, such as the \emph{IEEE/ACM International Conference on Program Comprehension} (ICPC) or the \emph{IEEE International Workshop on Source Code Analysis and Manipulation} (SCAM) that focus on source code comprehension and manipulation. 


We prune venues that may not contain relevant contributions to source code comments.  Specifically, we exclude a venue if its ten
years of proceedings contain fewer than five occurrences of the words
\emph{documentation} or \emph{comment}. 
This way, we exclude conferences, such as \emph{IEEE International Conference on Engineering of Complex Computer Systems}
(ICECCS), \emph{Foundations of Software Science and Computational
  Structures} (FoSSaCS), and many others that primarily focus on
other topics, such as verification or programming languages.
Thus, we reduce our dataset to 20 conferences and six journals, as shown in \tabref{venue-selection}.

In \tabref{venue-selection}, the column \emph{Type} specifies whether a venue is a conference (C) or a
journal (J), and the column \emph{Rank} denotes the corresponding CORE rank of the venue as of April 2021.
The column \emph{Selection} indicates the data collection phase in which the venue was first selected.
The column \emph{Papers per venue} indicates the total number of papers selected from this venue, both during the direct search and the snowball search.

We consider only full papers (published in a technical track and longer than five pages) since they are likely to be an extended or mature version of the papers published in other tracks, such as NIER, ERA, or Poster.

\subsubsection{Data Retrieval}
\label{subsec:data-retrieval}

We retrieve in step\,\ballnumber{2} the proceedings from January \asText\yearBegin to December \asText\yearEnd of the selected venues from the DBLP digital library. 
From each paper, we collect its metadata using the GitHub repository\footnote{https://github.com/sbaltes/dblp-retriever}, such as the title, authors, conference track (if present), its page length, and its Digital Object Identifier (DOI), directly from DBLP for a total of \asText\numPapersFomCrawl publications.
For each paper, the DOI is resolved and its abstract is collected from the publisher webpage.

\begin{table*}
  \footnotesize
  \caption{Included Journals, Conferences, and Workshops.}
  \tablabel{venue-selection}
  \begin{tabularx}{\linewidth}{llllll}\cline{1-6}
    \textbf{Venue} & \textbf{Abbreviation} & \textbf{Rank} & \textbf{Type}
    & \textbf{Selection} & \textbf{Papers per venue}
    \\\cline{1-6}
    \begin{filecontents*}{selected-venues.csv}
Venue,VenueAbbreviation,Rank,Type,Selection,Papers,
ACM Computing Surveys,CSUR,A*,J,search,,
ACM Transactions on Software Engineering and Methodology,TOSEM,A*,J,search,,
IEEE Transactions on Software Engineering,TSE,A*,J,search,5,
Empirical Software Engineering: an international journal,EMSE,A,J,search,6,
Journal of Systems and Software,JSS,A,J,search,2,
Information and Software Technology,IST,A,J,search,1,
ACM SIGSOFT Symposium on the Foundations of Software Engineering,ESEC/FSE,A*,C,search,3,
International Conference on Software Engineering,ICSE,A*,C,search,6,
Architectural Support for Programming Languages and Operating Systems,ASPLOS,A*,C,search,,
Computer Aided Verification,CAV,A*,C,search,,
International Conference on Functional Programming,ICFP,A*,C,search,,
ACM Conference on Object Oriented Programming Systems Languages and Applications,OOPSLA,A*,C,search,1,
ACM-SIGPLAN Conference on Programming Language Design and Implementation,PLDI,A*,C,search,,
ACM-SIGACT Symposium on Principles of Programming Languages ,POPL,A*,C,search,,
Measurement and Modeling of Computer Systems,SIGMETRICS,A*,C,search,,
Automated Software Engineering Conference,ASE,A,C,search,2,
International Conference on Evaluation and Assessment in Software Engineering,EASE,A,C,search,1,
International Symposium on Empirical Software Engineering and Measurement,ESEM,A,C,search,1,
IEEE International Conference on Software Maintenance and Evolution,ICSME,A,C,search,2,
IEEE International Working Conference on Mining Software Repositories,MSR,A,C,search,1,
International Symposium on Software Reliability Engineering,ISSRE,A,C,search,,
IEEE International Working Conference on Software Visualisation,VISSOFT,B,C,search,,
IEEE International Conference on Global Software Engineering,ICGSE,C,C,search,,
IEEE International Conference on Program Comprehension,ICPC,C,C,search,5,
International Workshop on Modelling in Software Engineering,MISE,C,C,search,,
IEEE International Workshop on Source Code Analysis and Manipulation,SCAM,C,C,search,1,
International Conference on Web Information Systems and Applications,WISA,-,C,snowball,1,
Software Quality Journal,-,C,J,snowball,1,
Software: Practice and Experience,SPE,B,J,snowball,1,
ACM Symposium on Applied Computing,SAC,B,C,snowball,1,
IEEE Workshop on Machine Learning Techniques for Software Quality Evaluation,MaLTeSQuE,-,C,snowball,1,
Journal of Software: Evolution and Process,JSEP,B,J,snowball,1,
Asia-Pacific Symposium on Internetware,Internetware,-,C,snowball,1,
IEEE International Joint Conference on Neural Networks,IJCNN,A,C,snowball,1,
International Computer Software and Applications Conference,COMPSAC,B,C,snowball,1,
Asia-Pacific Software Engineering Conference,APSEC,B,C,snowball,1,
International journal of software engineering and knowledge engineering,SEKE,-,J,snowball,1,
\end{filecontents*}
\csvreader[late after line=\\,
    late after last line=\\\cline{1-6}]{selected-venues.csv}%
    {Venue=\Venue, VenueAbbreviation=\VenueAbbreviation,Rank=\Rank,Type=\Type,Selection=\Selection,Papers=\Papers}%
    {\Venue & \VenueAbbreviation & \Rank & \Type & \Selection & \Papers}%
  \end{tabularx}
\end{table*}

\paragraph{Keyword-based filtering}
We apply in step\,\ballnumber{3} a keyword-based
search (given in \autoref{subsec:search-keywords}) using a prefix function to the retrieved proceedings to select potentially
relevant papers.
We account for possible upper- and lowercase letters in the keywords, and sometimes use variations of keywords (\eg singular and
plural forms). 

Our filtering will get papers (whose title and abstract include keywords from $K_{1}$ and $K_{2}$ but not from $K_{3}$) that explicitly mention concepts we are interested in, \eg \emph{``A Human Study of Comprehension and Code Summarization''} from ICPC 2020~\cite{Stap20a} is matched by keywords \emph{summar} from \emph{$K_{1}$} in the title and \emph{quality} from \emph{$K_{2}$} in the abstract, but will exclude papers not sufficiently close to our research subject, \eg \emph{``aComment: mining annotations from comments and code to detect interrupt related concurrency bugs''} from ICSE 2011 has two keywords \emph{comment} and \emph{annotation} from \emph{$K_{1}$} but none from the \emph{$K_{2}$}.

The final set of keywords we use for filtering is the result of an iterative approach: we manually scan the full venue proceedings metadata to make sure the set of keywords did not prune relevant papers, and we refine the set of keywords during several iterative discussions.
 This iterative approach gives us confidence that our keyword-based filtering approach does not lead to false negatives for the selected venues.
After applying the keyword-based filtering, we identify \asText\numPapersReviewedWithoutSnowball studies as potentially-relevant papers from a total of \asText\numPapersFomCrawl, which we review manually.

\subsubsection{Data selection}
\label{subsec:data-selection}
We analyze\,\ballnumber{4} the
\asText\numPapersReviewedWithoutSnowball selected papers following the protocol where four authors or evaluators manually evaluate the papers based on the inclusion and exclusion criterion to ensure that they indeed assess comment quality.

\noindent
\textbf{Inclusion criteria}
\begin{enumerate}[label=I\arabic*,start=1]
\item The topic of the paper is about code comment quality.
\item The study presents a model/technique/approach to assess code
  comments or software documentation including code comments.
\end{enumerate}

\noindent
\textbf{Exclusion criteria}
\begin{enumerate}[label=E\arabic*,start=1]
  \item The paper is not in English.
  \item It does not assess any form of quality aspects of comments \eg
    content, style, or language used.
  \item It is not published in a technical track.
  \item It is a survey paper.
  \item It is not a peer reviewed paper, or it is a pre-print.
  \item It covers other documentation artifacts, \ie not comments.
  \item It is shorter than 5 pages.
\end{enumerate}

\paragraph{Manual analysis}

The selected papers were equally divided among four evaluators (\ie two Ph.D. candidates and two faculty members) based on years of publications so that each evaluator gets papers from all venues, \eg the first author evaluate proceedings from 2011 to 2013. 
We make sure that evaluators do not take decisions on papers they co-authored to avoid conflicts of interest.
Each evaluator has at least two years of experience in the domain of comment analysis.
Each paper is reviewed by three evaluators.
The evaluators follow a three-iteration-based process to evaluate the assigned papers.
In the first iteration, the first evaluator independently assesses the relevance of a paper based on the criteria by inspecting each paper's
title and abstract, to make an initial guess, then inspecting its conclusion to reach the final decision.
In the next iteration, another evaluator reviews the paper and validates the previous decision by adding the label ``agrees/disagrees with the first evaluator''. 
With this process, every publication selected in the final set is reviewed by at least two researchers. In case they do not agree, the third evaluator reviews it~\cite{Kuhr17a}, and the final decision is taken based on the majority voting mechanism.

We decide, for instance, to include the study by Hata \etal~\cite{Hata19a}, even though it only talks about links in comments. 
 Though it does not explicitly describe any quality aspect of comments, it mentions the traceability of the links, which is a \qa we consider in our study.
All studies considered in our SLR together with their evaluation (the agreement and disagreement for each study) are available in our
replication package. 

Thus, we reduce \asText\numPapersReviewedWithoutSnowball
papers to \asText\numPapersCandidateWithoutSnowball
candidate papers (\ie
\asText{\FPuse{round(numPapersCandidateWithoutSnowball/\numPapersReviewed*100,
    0)}\%}) with a fair agreement according to Cohen's Kappa (k=0.36).
For all candidate papers, we read in step\,\ballnumber{5} their introduction, conclusion, and the study design (if needed), and discuss them amongst ourselves to ensure their relevance.  During this analysis process, some
additional papers were found to be irrelevant.  For example, the study
by Aghajani \etal seems relevant based on the title and abstract, but
does not really evaluate code comments, and we thus
discarded it~\cite{Agha20b}.  With this process,
\asText{\FPuse{round(\numPapersCandidateWithoutSnowball-\numPapersRelevantWithoutSnowball, 0)}} papers in total were
discarded, reducing the relevant paper set to
\asText\numPapersRelevantWithoutSnowball papers.

\paragraph{{Data gathering for snowballing}}
To include further relevant papers that we might have missed with the venue-based approach, we perform in step\,\ballnumber{6} a forward and backward
snowballing approach for the \asText\numPapersRelevantWithoutSnowball papers and retrieve a total of
\asText{\FPuse{round(\numSnowballCitationsUnique+\numSnowballRefsUnique,0)}} unique papers.

\begin{center}
\begin{tabular}{l|rrr}
Snowball papers & Total & Unique & \emph{Selected} 
\\ \hline
from citations
& \asText{\numSnowballCitationsTotal}
& \asText{\numSnowballCitationsUnique}
& \asText{\numSnowballCitationsSelected}
\\
from references
& \asText{\numSnowballRefsTotal}
& \asText{\numSnowballRefsUnique}
& \asText{\numSnowballRefsSelected}
\\
\end{tabular}
\end{center}

The column \emph{Total} reports the total number of references and citations
collected. The \emph{Unique} column reports a total number of unique items (\ie since relevant papers cover similar topics many references, and citations are shared across our set of studies). 
Finally, the column \emph{Selected} reports the total number of unique references and citations whose publication year falls within our time frame range, \ie 2011-2020.

\paragraph{Data selection from snowballing}

We repeat in step \ballnumber{7} the same keyword-based filtering to these \asText{\FPuse{round(\numSnowballCitationsUnique+\numSnowballRefsUnique,0)}} papers, as described in~\autoref{subsec:autom-data-coll}. As a result,
\asText\numPapersSnowballAddedAll papers were added for manual analysis.  We repeat in step\,\ballnumber{8} the three-iteration based manual analysis process and find 
\asText\numPapersSnowballSelected
additional candidate papers to analyze.
After the second round of discussion\,\ballnumber{9} we keep \asText\numPapersSnowballRelevant additional relevant papers.
We find a total of \asText\numPapersRelevant papers shown in \tabref{included-studies} published in the venues shown in \tabref{venue-selection}. In \tabref{included-studies}, the column \emph{Study ID} indicates the ID assigned to each paper, the column \emph{Title} presents the title of the paper, and the column \emph{Year} indicates the years in which the paper is published.

To further ensure the relevance of our search strategy, we search our keywords on popular publication databases, such as ACM, IEEE Xplore, Wiley \etc 
We search for our keywords in titles and abstracts.\footnote{It is not possible to search the keywords in abstracts in Wiley.}
We retrieve 13\,144 results from IEEE Xplore, and 10\,567 from ACM for the same timeline (\yearBegin-\yearEnd).
We inspect first 200 results (sorted by relevance criterion on the publisher webpage) from each of these databases. We apply our inclusion and exclusion criterion to find the extent to which our venue selection criteria might have missed relevant papers.
Our results from ACM show that 19\% of the these papers are already covered by our search strategy but only 5\% of them fulfilled our inclusion criterion.
Nearly 81\% of the papers are excluded due to their non-SE venue.
Among these papers, 80\% are unrelated to the code comment quality aspect while 1\% of papers (two papers) that are related to code comments are missed due to two main reasons, (i) the venue not being indexed in CORE2020, and (ii) the paper being from a non-technical track.
Similarly, the results from IEEE show that 30\% of the papers are already covered by our search strategy but only 5\% of them fulfilled the inclusion criterion. Nearly 69\% of the papers are excluded due to their non-SE venue and unrelated to code comment quality aspect. 
We also find 1\% papers that are relevant to our topic of interest but excluded due to the length criteria, specifically one of the paper is a poster paper and another is a short paper.

\begin{center}
\begin{table*}
  \footnotesize
  \renewcommand{\arraystretch}{0.96}
  \caption{Included studies}
  \tablabel{included-studies}
  \begin{tabularx}{\linewidth}{llll}
  \cline{1-4}
    \textbf{Study ID} & \textbf{Title} & \textbf{Year} & \textbf{Reference} \\
  \cline{1-4}
  \begin{filecontents*}{papers.csv}
sid,Paper ID,title,year,reference,reference_cite
S01,1,How Good is Your Comment? A Study of Comments in Java Programs.,2011,Haou11a,\cite{Haou11a}
S02,5,Quality Analysis of Source Code comments.,2013,Stei13b,\cite{Stei13b}
S03,7,Evaluating Usage and Quality of Technical Software Documentation: An Empirical Study.,2013,Garo13a,\cite{Garo13a}
S04,10,Inferring Method Specifications from Natural Language API Descriptions.,2012,Pand12a,\cite{Pand12a}
S05,16,Using Traceability Links to Recommend Adaptive Changes for Documentation Evolution.,2014,Dage14a,\cite{Dage14a}
S06,33,On Using Machine Learning to Identify Knowledge in API Reference Documentation.,2019,Fucc19a,\cite{Fucc19a}
S07,20,Detecting Fragile Comments.,2017,Rato17a,\cite{Rato17a}
S08,21,Automatically Assessing Code Understandability: How Far are We?,2017,Scal17a,\cite{Scal17a}
S09,23,Analyzing APIs Documentation and Code to Detect Directive Defects.,2017,Zhou17a,\cite{Zhou17a}
S10,26,The Effect of Poor Source Code Lexicon and Readability on Developers' Cognitive Load.,2018,Fakh18a,\cite{Fakh18a}
S11,28,A Large-Scale Empirical Study on Linguistic Antipatterns Affecting APIs.,2018,Agha18a,\cite{Agha18a}
S12,29,Improving API Caveats Accessibility by Mining API Caveats Knowledge Graph.,2018,Lili18a,\cite{Lili18a}
S13,32,A Learning-Based Approach for Automatic Construction of Domain Glossary from Source Code and Documentation.,2019,Wang19b,\cite{Wang19b}
S14,35,A Framework for Writing Trigger-Action Todo Comments in Executable Format.,2019,Nie19a,\cite{Nie19a}
S15,36,A Large-Scale Empirical Study on Code-Comment Inconsistencies.,2019,Wen19a,\cite{Wen19a}
S16,38,Software Documentation Issues Unveiled.,2019,Agha19a,\cite{Agha19a}
S17,42,The Secret Life of Commented-Out Source Code.,2020,Pham020,\cite{Pham020}
S18,88,Code Comment Quality Analysis and Improvement Recommendation: An Automated Approach,2016,Sun16a,\cite{Sun16a}
S19,45,A Human Study of Comprehension and Code Summarization.,2020,Stap20a,\cite{Stap20a}
S20,46,CPC: Automatically Classifying and Propagating Natural Language Comments via Program Analysis.,2020,Zhai20a,\cite{Zhai20a}
S21,48,Recommending Insightful Comments for Source Code using Crowdsourced Knowledge.,2015,Rahm15b,\cite{Rahm15b}
S22,49,Improving Code Readability Models with Textual Features.,2016,Scal16a,\cite{Scal16a}
S23,54,Automatic Source Code Summarization of Context for Java Methods.,2016,McBu16a,\cite{McBu16a}
S24,55,Automatic Detection and Repair Recommendation of Directive Defects in Java API Documentation.,2020,Zhou20a,\cite{Zhou20a}
S25,58,Measuring Program Comprehension: A Large-Scale Field Study with Professionals.,2018,Xia18a,\cite{Xia18a}
S26,62,Usage and Usefulness of Technical Software Documentation: An Industrial Case Study,2015,Garo15a,\cite{Garo15a}
S27,64,What Should Developers be Aware of? An Empirical Study on the Directives of API Documentation.,2012,Monp12a,\cite{Monp12a}
S28,65,Analysis of License Inconsistency in Large Collections of Open Source Projects.,2017,Wu17a,\cite{Wu17a}
S29,70,Classifying Code Comments in Java Software Systems.,2019,Pasc19a,\cite{Pasc19a}
S30,71,Augmenting Java Method Comments Generation with Context Information based on Neural Networks.,2019,Zhou19a,\cite{Zhou19a}
S31,74,Improving Source Code Lexicon via Traceability and Information Retrieval.,2011,Luci11a,\cite{Luci11a}
S32,77,Detecting API Documentation Errors.,2013,Zhon13a,\cite{Zhon13a}
S33,99,Analyzing Code Comments to Boost Program Comprehension,2018,Shin18a,\cite{Shin18a}
S34,85,Recommending Reference API Documentation.,2015,Robi15a,\cite{Robi15a}
S35,86,Some Structural Measures of API Usability,2015,Rama15a,\cite{Rama15a}
S36,87,An Empirical Study of the Textual Similarity between Source Code and Source Code Summaries.,2016,McBu16b,\cite{McBu16b}
S37,90,Linguistic Antipatterns: What They are and How Developers Perceive Them.,2016,Arna16a,\cite{Arna16a}
S38,91,Coherence of Comments and Method Implementations: A Dataset and An Empirical Investigation,2016,Cora18a,\cite{Cora18a}
S39,97,A Comprehensive Model for Code Readability,2018,Scal18a,\cite{Scal18a}
S40,101,Automatic Detection of Outdated Comments During Code Changes,2018,Liu18a,\cite{Liu18a}
S41,103,Classifying Python Code Comments Based on Supervised Learning,2018,Zhan18a,\cite{Zhan18a}
S42,106,Investigating Type Declaration Mismatches in Python,2018,Pasc18a,\cite{Pasc18a}
S43,108,The Exception Handling Riddle: An Empirical Study on the Android API.,2018,Kech18a,\cite{Kech18a}
S44,109,"9.6 Million Links in Source Code Comments: Purpose, Evolution, and Decay.",2019,Hata19a,\cite{Hata19a}
S45,115,Migrating Deprecated API to Documented Replacement: Patterns and Tool,2019,Xi19a,\cite{Xi19a}
S46,120,A Topic Modeling Approach To Evaluate The Comments Consistency To Source Code,2020,Iamm20a,\cite{Iamm20a}
S47,123,Comparing Identifiers and Comments in Engineered and Non-Engineered Code: A Large-Scale Empirical Study,2020,Lemo20a,\cite{Lemo20a}
\end{filecontents*}
\csvreader[late after line=\\,
  late after last line=\\\cline{1-4}]{papers.csv}%
  {sid=\sid, title=\title,year=\year,reference_cite=\reference}
  {\sid & \title & \year & \reference}%
  \end{tabularx}
\end{table*}
\vspace{-1mm}
\end{center}

\subsubsection{Data Evaluation}
\label{subsec:data-evaluation}
We work in step\,\ballnumber{10} on the full versions of the \asText\numPapersRelevant relevant
papers to identify the \qas and the approaches to assess comments.  
In case we cannot retrieve the full PDF version of a paper, we use university resources to access it.
This affects only one paper by Sun~\etal, which requires payment to access the full version~\cite{Sun16a}.
In case we cannot access a paper via any resource, we remove it from our list. We find no such inaccessible study.

We report all papers in an online shared spreadsheet on Google Drive to facilitate their analysis collaboratively.
For each paper we extract common metadata, namely \emph{Publication year}, \emph{Venue}, \emph{Title}, \emph{Authors}, \emph{Authors' country}, and \emph{Authors' affiliation}. 
We then extract various dimensions (described in the following paragraphs) formulated to answer all research questions.

\subsection{{Data extraction for research questions}}
\seclabel{data-extraction-research-ques tions}
To answer \emph{RQ1} {(\emph{\rqI})}, we record the \emph{Comment scope} dimension. It lists the scope of comments under assessment such as class, API, method (function), package, license, or inline comments. In case the comment type is not mentioned, we classify it as ``code comments''.
Additionally, we identify the programming languages whose comments are analyzed, and record this in the \emph{Language analyzed} dimension.

  \begin{table*}
    \centering
    \footnotesize
    \caption{RQ2 \qas mentioned by Zhi \etal (highlighted in bold) and other works}
    \tablabel{paper-fields-extraction-rq}
    \begin{tabular}
    {p{0.21\linewidth}p{0.19\linewidth}p{0.49\linewidth}}
    \noalign{\smallskip}\hline
    \textbf{Quality Attribute (\qa)} & \textbf{Synonyms} &  \textbf{Description}  \\
    \noalign{\smallskip}\hline
    \textit{\qas mentioned by Zhi \etal} \\
    \cline{1-1}
    \textbf{Accessibility}       &  \textbf{availability}, \textbf{information hiding}, \textbf{easiness to find} &      whether comment content can be accessed or retrieved by developers or not   \\
    \cline{3-3}
    \textbf{Readability}         & \textbf{clarity} &      the extent to which comments can be easily read by other readers         \\
    \cline{2-3}
    \textbf{Spelling and grammar}& natural language quality &      grammatical aspect of the comment content \\
    \cline{2-3}
    \textbf{Trustworthiness}     & &      the extent to which developers perceive the comment as trustworthy   \\
    \cline{2-3}
    \textbf{Author-related}       & &     identity of the author who wrote the comment  \\
    \cline{2-3}
    \textbf{Correctness}         & &     whether the information in the comment is correct or not \\
    \cline{2-3}
    \textbf{Completeness}        & adequacy &     how complete the comment content is to support development and maintenance tasks or whether there is missing information in comments or not \\
    \cline{2-3}
    \textbf{Similarity}          & \textbf{uniqueness}, \textbf{duplication} &    how similar the comment is to other code documents or code\\
    \cline{2-3}
    \textbf{Consistency}         & \textbf{uniformity}, \textbf{integrity} &   the extent to which the comment content is consistent with other documents or code \\   
    \cline{2-3}
    \textbf{Traceability}        & &   the extent to which any modification in the comment can be traced, including who performed it                          \\
    \cline{2-3}
    \textbf{Up-to-datedness}     & &   how the comment is kept up-to-date with software evolution                            \\
    \cline{2-3}
    \textbf{Accuracy}            & \textbf{preciseness} & accuracy or preciseness of the comment content. If the documentation is too abstract or vague and does not present concrete examples, then it can seem imprecise.                      \\
    \cline{2-3}
    \textbf{Information organization} & &     how the information inside a comment is organized in comments                       \\
    \cline{2-3}
    \textbf{Format} & including visual models, use of examples &    quality of documents in terms of writing style, description perspective, use of diagrams or examples, spatial arrangement, etc.                       \\
    \noalign{\smallskip}\hline
    \textit{\qas mentioned by other works} \\
    \cline{1-1}
    Coherence & &  how comment and code are related to each other, \eg method comment should be related to the method name(S02, S38) \\
    \cline{2-3}
    Conciseness & & the extent to which comments are not verbose and do not contain unnecessary information (S23, S30, S47)  \\
    \cline{2-3}
    Content relevance & & how relevant the comment or part of the comment content is to a particular purpose (documentation, communication) (S01, S03, S29, S41, S47) \\
    \cline{2-3}
    Maintainability & & the extent to which comments are maintainable (S15-S17,S20-S21)  \\
    \cline{2-3}
    Understandability & & the extent to which comments contribute to understanding the system (S19, S23) \\
    \cline{2-3}
    Usability & Usefulness &  to which extent the comment can be used by readers to achieve their objectives (S02, S16, S34, S35)\\
    \cline{2-3}
    Documentation technology &   & whether the technology to write, generate, store documentation is current or not\\
    \cline{2-3}
    Internationalization & & the extent to which comments are correctly translated in other languages (S16) \\
    \cline{2-3}
    Other  & &  the study does not mention any \qa and cannot be mapped to any of the above attributes                       \\
    \noalign{\smallskip}\hline
    \end{tabular}
  \end{table*}

To answer \emph{RQ2} {(\emph{\rqII})}, we identify various \qas researchers mention to assess comment quality.
This reflects the various quality aspects researchers perceive as important to have high-quality comments.
\tabref{paper-fields-extraction-rq} lists the \qas in the \emph{Quality attribute (\qa)} column and their brief summary in the 
\emph{Description} column.
 Of these \qas, several are mentioned by Zhi \etal in their work~\cite{Zhi15a}, and are highlighted by the bold text compared to \qas mentioned in other works.
As Zhi \etal considered various types of documentation, such as requirement and architectural documents, not all attributes fit exactly into our study.
For instance, the category ``Format'' includes the format of the documentation (\eg UML, flow chart) in addition to the other aspects such as writing style of the document, use of diagrams \etc
Although the format of the documentation is not applicable in our case due to our comment-specific interest, we keep other applicable aspects (writing style, use of diagram) of this \qa.
In addition to their \qas, we include any additional attribute mentioned in our set of relevant papers.  
If a study uses different terminology but similar meaning to \qas in our list, we map such \qas to our list and update the list of possible synonyms as shown in the column \emph{Synonyms} in \tabref{paper-fields-extraction-rq}.
In case we cannot map a study to the existing \qas, we map it to the \emph{Other} category.

For the cases where the studies do not mention any specific \qa and mention
comment quality analysis in general, we map the study to the list of existing \qas or classify it as \emph{Other} based  on their goal behind the quality analysis. 
For example, Pascarella~\etal identify various information types in comments to support developers in easily finding relevant information for code comprehension tasks and to improve the comment quality assessment~\cite{Pasc17a}.
They do not mention any specific \qa, but based on their study goal of finding relevant information easily, we map their study to the \emph{content relevance} \qa.
Similarly, we map other comment classification studies such as S06, S29, S33, and S41 to the \emph{content relevance} attribute.
At the same time, the studies on linguistic anti-patterns (LAs) are mapped to the \emph{consistency} attribute, given that LAs are practices that lead to lexical inconsistencies among code elements, or between code and associated comments~\cite{Arna16a,Fakh18a,Agha18a}.
Additionally, the studies that mention the negation of the \qas such as \emph{inconsistency}, \emph{incorrectness}, or \emph{incompleteness} are mapped to their antonyms as \emph{consistency}, \emph{correctness}, or \emph{completeness}, respectively to prevent duplication.

\emph{RQ3} {(\emph{\rqIII})} concerns various methods researchers use or propose to assess comment \qas, for instance, whether they use machine-learning based methods to assess comment quality or not.
\begin{itemize}
\item \emph{Technique type.}  This identifies whether the technique used to assess a \qa is based on natural language processing (NLP), heuristics, static analysis, metrics, machine-learning (ML), or deep neural network (DNN) approaches. The rationale is to identify which \qas are often assessed manually or using a specific automated approach.
For instance, if the study uses specific heuristics related to the programming environment to assess a \qa, it is classified as \emph{heuristic-based} technique, if it uses abstract syntax tree (AST) based static analysis approaches, then it is assigned to \emph{static analysis}, and if it uses machine-learning or deep-learning-based techniques (including any or both of the supervised or unsupervised learning algorithms), then it is classified as \emph{ML-based}, or \emph{DNN-based} respectively.
A study can use mixed techniques to assess a specific \qa and thus can be assigned to multiple techniques for the corresponding \qa.
We often find cases where the studies do not use any automated technique to measure a \qa and instead ask other developers to assess it manually, so we put such cases into the \emph{manual assessment} category.
In case the study mentions a different technique, we extend the dimension values.

\item \emph{Metrics or tools.} This further elaborates specific metrics, or tools the studies propose or use to assess a \qa. A study can use an existing metric or can propose a new one. Similarly, one metric can be used to assess multiple \qas. We identify such metrics to highlight popular metrics amongst researchers.
\end{itemize}

 \emph{RQ4} {(\emph{\rqIV})} captures the nature of the study and the type of contribution researchers use or propose to assess comment quality. 
 We first identify the nature of research of a study and then identify the type of contribution it provides.
 This can reflect the kind of research often conducted to assess comment quality and the kind of contribution they make to support developers in assessing comment quality, for instance, what kind of solutions the \emph{Solution Proposal} research often propose, such as a method, metric, model, or tool. 
 
 \begin{table*}
    \centering
    \footnotesize
    \caption{Type of research approach studies use and type of contributions studies make}
    \tablabel{paper-research-contribution-type}
    \begin{tabular}
    {p{0.12\linewidth}p{0.15\linewidth}p{0.69\linewidth}}
    \noalign{\smallskip}\hline
    \textbf{Dimension}  & \textbf{Category} & \textbf{Description}  \\
    
    \noalign{\smallskip}\hline
    	\textbf{Research type}  & Empirical  & This research task focuses on understanding and highlighting various problems by analyzing relevant projects, or surveying developers. These papers often provide empirical insights rather than a concrete technique. \\
	\noalign{\smallskip}\cline{2-3}
    
	  & Validation  & This research task focus on investigating the properties of a technique that is novel and is not yet implemented in practice, \eg techniques used for mathematical analysis or lab experimentation \\
	\noalign{\smallskip}\cline{2-3}
	    
	& Evaluation & The paper investigates the techniques that are implemented in practice and their evaluation is conducted to show the results of the implementation in terms of its pros and cons and thus help researchers in improving the technique. \\
	\noalign{\smallskip}\cline{2-3}
    
	& Solution Proposal & The paper proposes a novel or a significant extension of an existing technique for a problem and describes its applicability, intended use, components, and how the components fit together using a small example or argumentation. \\
	\noalign{\smallskip}\cline{2-3}
    
	& Philosophical  &  These papers present a new view to look at the existing problems by proposing a taxonomy or a conceptual framework, \eg developing a new language or framework to describe the observations is a philosophical activity.\\
	\noalign{\smallskip}\cline{2-3}

	& Opinion & These papers describe the author's opinion in terms of how things should be done, or if a certain technique is good or bad. They do not rely on research methodologies and related work.\\
	\noalign{\smallskip}\cline{2-3}
    
	& Experience & These papers explain the personal experience of a practitioner in using a certain technique to show how something has been done in practice. They do not propose a new technique and are not scientific experiments.\\
	
        \noalign{\smallskip}\hline
	 \textbf{Contribution type}  & Empirical  & The paper provides empirical results based on analyzing relevant projects to understand and highlights the problems related to comment quality.  \\
	\noalign{\smallskip}\cline{2-3}
    
	& Method/technique  & The paper provides a novel or significant extension of an existing approach. \\
	\noalign{\smallskip}\cline{2-3}
    
	& Model  &  Provides a taxonomy to describe their observations or an automated model based on machine/deep learning. \\
	\noalign{\smallskip}\cline{2-3}

	& Metric & Provides a new metric to assess specific aspects of comments. \\
	\noalign{\smallskip}\cline{2-3}
	
	& Survey & Conducts survey to understand a specific problem and contribute insights from developers. \\
	\noalign{\smallskip}\cline{2-3}
	
	& Tool & Develops a tool to analyze comments. \\
	\noalign{\smallskip}\hline
	
    \end{tabular}
  \end{table*}
 
To capture this information, we formulate the following dimensions: 
\begin{itemize}
\item \emph{Research type.} This identifies the nature of the research approach used in the studies, such as empirical, validation, evaluation, solution proposal, philosophical, opinion, or experience paper~\cite{Wier06a,Pete08a}.
The dimension values are described in detail in \tabref{paper-research-contribution-type}.

\item \emph{Paper contribution.} This dimension describes the type of
contribution the study provides in terms of a method/technique,
tool, process, model, metric, survey, or empirical results~\cite{Pete08a}. 
The dimension values are described in detail in \tabref{paper-research-contribution-type}.
If we cannot categorize it into any of these, we mark it ``Other''.

\item \emph{Tool availability.} This reflects whether the tool proposed in the study is accessible or not at the time of conducting our study. Gonz{\'a}lez \etal identified the reproducibility aspects characterizing empirical software engineering studies~\cite{Gonz12a} in which availability of the artifact (the tool proposed in the study, or the dataset used to conduct the study) is shown as an important aspect to facilitate the replication and extension of the study. Therefore, we record the availability of the proposed tool in this dimension and the availability of the dataset in the following dimension.

\item \emph{Dataset availability.} This reflects if the dataset used in the empirical study is accessible or not.
\end{itemize}

\emph{RQ5} {(\emph{\rqV})} concerns how various kinds of research (\emph{Research type} dimension described in the previous RQ), and various kinds of contribution (\emph{Paper contribution} dimension) are evaluated in the studies.
For example, it helps us to observe that if a study proposes a new method/technique to assess comments, then the authors also conduct an experiment on open-source projects to validate the contribution, or they consult the project developers, or both.
We capture the type of evaluation in the \emph{Evaluation type} dimension, and its purpose in \emph{Evaluation purpose}. 
The rationale behind capturing this information is to identify the shortcomings in their evaluations, \eg how often the studies proposing a tool are validated with practitioners.
\begin{itemize}

\item \emph{Evaluation type.} 
 It states the type of evaluation the studies conduct to validate their approaches, such as conducting an experiment on open-source projects (\emph{Experiment}), or surveying students, practitioners, or both. For the automated approaches, we consider various performance metrics, also known as Information Retrieval (IR) metrics, that are used to assess the machine/deep learning-based models, such as Precision, Recall, F1 Measure, or Accuracy under the \emph{performance metrics}.
In case the approach is validated by the authors of the work, we identify the evaluation type as \emph{Authors of the work}.

\item \emph{Evaluation purpose.} It states the motivation of evaluation by authors such as evaluate the functionality, efficiency, applicability, usability, accuracy, comment quality in general, or importance of attributes.
\end{itemize}

  
\section{Results}
\seclabel{results}

As mentioned in \autoref{subsec:data-selection}, we analyze
\asText\numPapersRelevant relevant papers in total. Before answering
our four RQs, we present a brief overview of the metadata (publishing venues) of the papers.

\begin{figure}[t]
  \centering
  \includegraphics[width=0.4\textwidth]{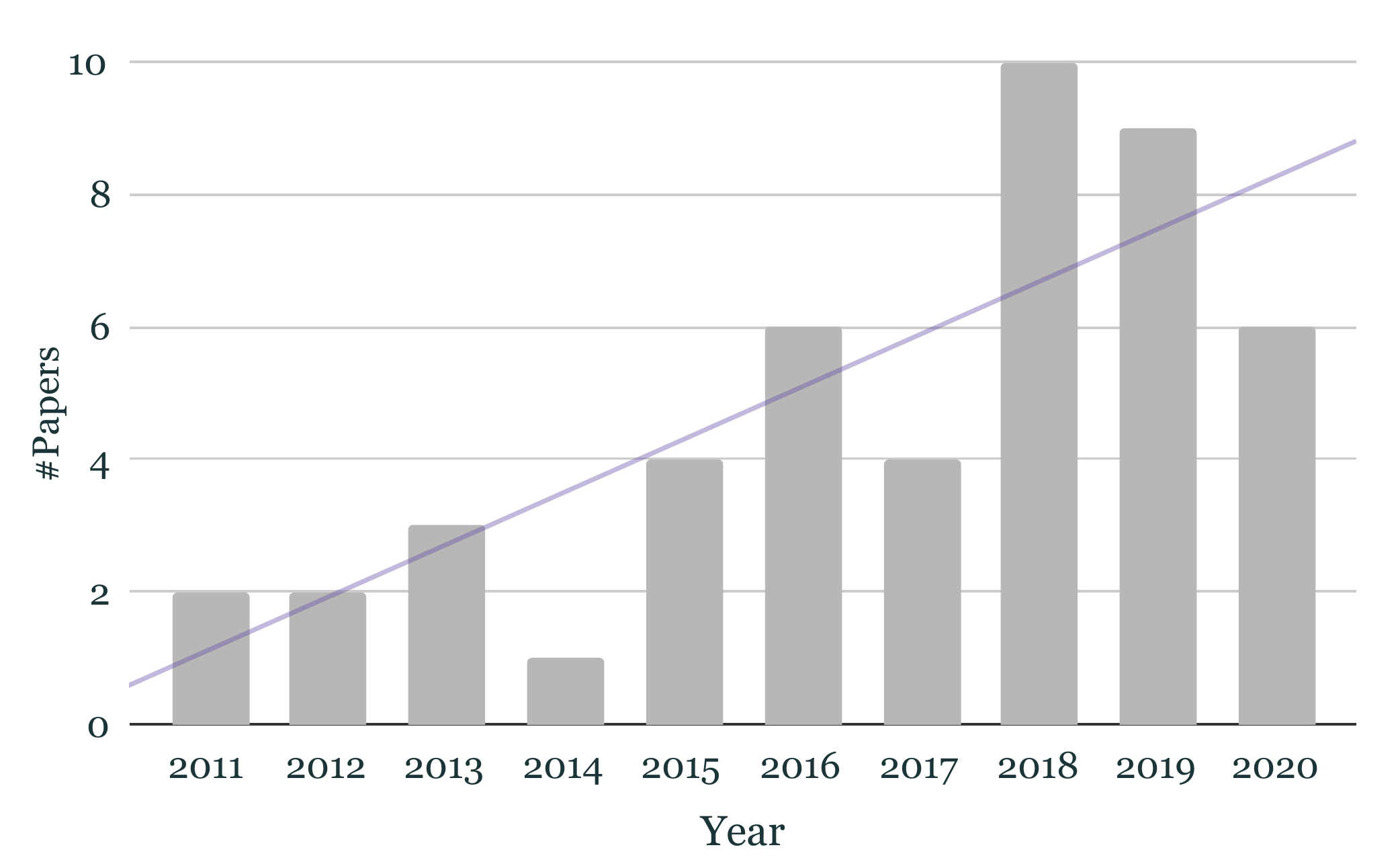}
  \caption{Relevant papers by years}
  \figlabel{plot-papers-by-years}
\end{figure}

\begin{figure*}
	\centering
	\begin{subfigure}[b]{0.40\textwidth}
		\centering
		\includegraphics[width=\textwidth]{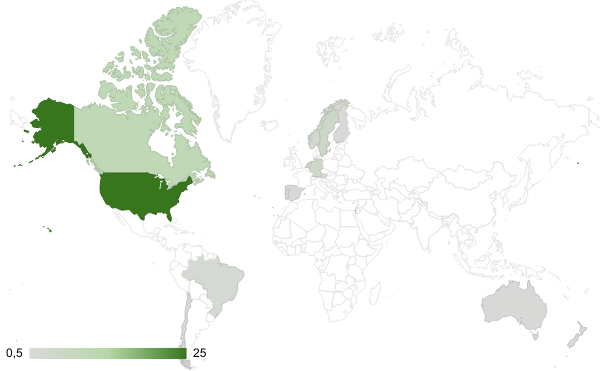}
		\caption{Zhi \etal \cite{Zhi15a} 1971-2011, all countries}
	\end{subfigure}
	\hfill
	\begin{subfigure}[b]{0.40\textwidth}
		\centering
		\includegraphics[width=\textwidth]{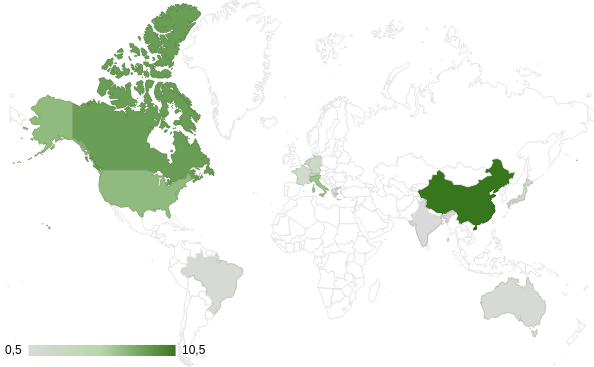}
		\caption{Our work 2011-2021, all countries}
	\end{subfigure}\\

  \begin{subfigure}[b]{0.40\textwidth}
		\centering
		\includegraphics[width=\textwidth]{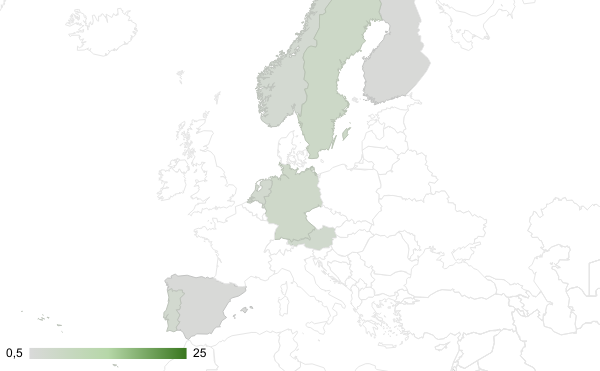}
		\caption{Zhi \etal \cite{Zhi15a} 1971-2011, Europe only}
	\end{subfigure}
	\hfill
	\begin{subfigure}[b]{0.40\textwidth}
		\centering
		\includegraphics[width=\textwidth]{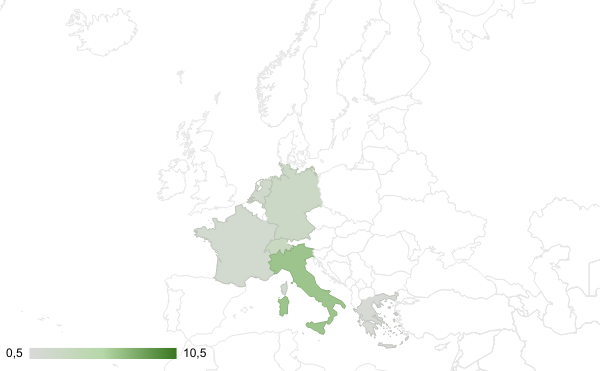}
		\caption{Our work 2011-2021, Europe only}
	\end{subfigure}
    \caption{Relevant papers by countries}
  \figlabel{plot-papers-by-countries}
\end{figure*}

\tabref{venue-selection} highlights the publication venues of
these papers. Most studies were published in top-tier software
engineering conferences (\eg ICSE) and journals, especially the ones
with a focus on empirical studies (\eg EMSE). This means that the
SE community agrees that assessing comment quality is an important topic deserving of research effort.
\figref{plot-papers-by-years} shows the paper distribution over the past decade, indicating a clear trend of increasing interest of the SE research community in comment quality assessment.
\figref{plot-papers-by-countries} shows the author distribution of the selected papers by the institution.
For the timeline 1971-2011, we rely on the geographical statistics data from the replication package of our reference study by Zhi~\etal~\cite{Zhi15a}, while for the period 2011-2021, and we collect these statistics as follows.
For each paper, the primary affiliations of all authors are taken into account.
If people from different countries co-authored a paper, we calculate the proportion of a country's contribution for each paper so that each paper gets a total score of one to avoid over-representing papers. For example, if five authors of a paper belong to Switzerland and one belongs to Spain, we assign 5/6 score for Switzerland and 1/6 for Spain for the paper.
Comparison with the previous data allows us to see the evolution of the field, with more even distribution of researchers nowadays and (unsurprising) rise of contributions from southeast Asia, specifically from China.

\finding{1}{The trend of analyzing comment quality has increased in the last decade (2011-2020), in part due to more researchers from southeast Asia working on the topic.}


\subsection{RQ$_1$: \rqI}
\seclabel{resultsRq1}

\begin{figure}[t]
	\centering
	\includegraphics[width=0.50\textwidth]{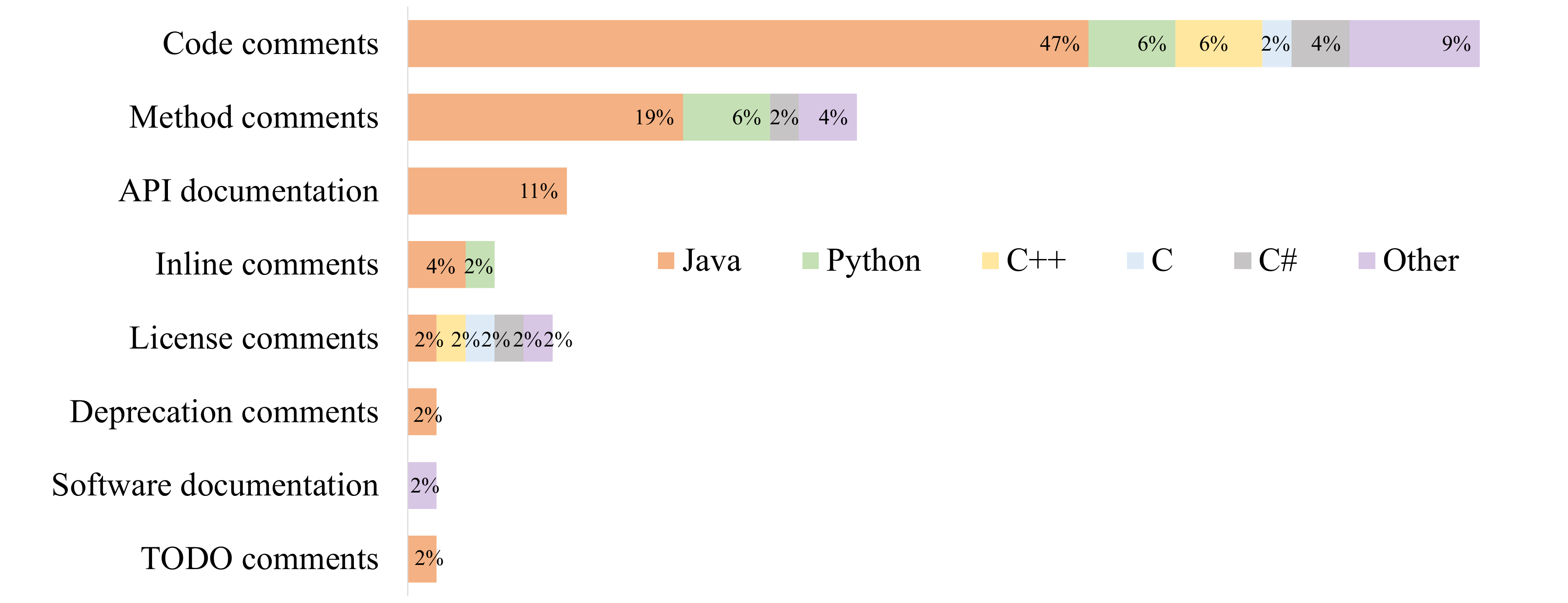}
	\caption{Types of comments per programming language}
	\figlabel{plot-comments-by-languages}
\end{figure}

To describe the rationale behind code implementation, various programming languages use source code comments. Our results show that researchers focus more on some programming languages compared to others as shown in 
\figref{plot-comments-by-languages}. 
This plot highlights the types of comments on the y-axis; each stack in the bar shows the ratio of the studies belonging to a particular language.
For instance, the majority (87\%) of the studies focus on code comments from Java, whereas only 15\% of the studies focus on code comments from Python, and 10\% of them focus on C\# and C$++$.
These results are in contrast to popular languages indicated by various developer boards, such as GitHub, Stack Overflow, or TIOBE. For instance, the TIOBE index show Python and C languages more popular than Java.\footnote{\url{https://www.tiobe.com/tiobe-index/} verified on Sep, 2021}
Similarly, the developer survey of 2019 and 2020 by Stack Overflow show that Java stands fifth after JavaScript, HTML/CSS, SQL, and Python among the most commonly used programming languages.\footnote{\url{https://insights.stackoverflow.com/survey/2020}}
We find only one study (S44) that seems to address the comment quality aspect in JavaScript.
Given the emerging trend of studies leveraging natural-language information in JavaScript code~\cite{Motw19a, Mali19a}, more research about comment quality may be needed in this environment.
It indicates that researchers need to analyze comments of other languages to verify their proposed approaches and support developers of other languages.

\finding{2}{87\% of the studies analyze comments from Java while other languages have not yet received enough attention from the research community.}

As code comments play an important role in describing the rationale behind source code, various programming languages use different types of comments to describe code at various abstraction levels.
For example, Java class comments should present high-level information about the class, while method comments should present implementation-level details~\cite{Nurv03a}. 
We find that half of the studies (51\% of the studies) focus on all types of comments whereas the other half focus on specific types of comments, such as inline, method, or TODO comments.
However, we also see in \figref{plot-comments-by-languages} that studies frequently focus on method comments and API documentation.
This proves the effort the research community is putting into improving API quality. 
While some attention is given to often overlooked kinds of comments, such as license comments (S28,S33), TODO comments (S14), inline comments (S17), and deprecation comments (S45), no relevant paper seems to focus specifically on the quality of \textit{class} or \textit{package} comments. 
Recently Rani \etal studied the characteristics of class comments of Smalltalk in the Pharo environment\footnote{\url{https://pharo.org/}} and highlighted the contexts they differ from Java and Python class comments, and why the existing approaches (based on Java, or Python) need heavy adaption for Smalltalk comments~\cite{Rani21b,Rani21d}.
This may encourage more research in that direction, possibly for other programming languages.

\finding{3}{Even though 50\% of the studies analyze all types of code comments, the rest focus on studying a specific type of comments such as method comments, or API comments, indicating research interest in leveraging a particular type of comment for specific development tasks.}

Previous work by Zhi \etal showed that a majority of studies analyze just one type of system~\cite{Zhi15a}. 
In contrast, our findings suggest that the trend of analyzing comments of multiple languages and systems is increasing. 
For example, 80\% of the studies analyzing comments from Python and all studies analyzing comments from C$++$ also analyze comments from Java.
Only Pascarella \etal (S42) and Zhang \etal (S41) focus solely on Python~\cite{Pasc18a,Zhan18a}.
However, Zhang \etal (S41) perform the comment analysis work in Python based on the Java study (S29) by Pascarella \etal~\cite{Zhan18a,Pasc17a}.
Such trends also reflect the increasing use of polyglot environments in software
development~\cite{Toma14a}. The ``Other'' label in \figref{plot-comments-by-languages} comprises
language-agnostic studies, \eg S16 
 or the studies considering less popular languages,
\eg S28 
focuses on COBOL. 
We find only one study (S44) that analyzes comments of six programming languages \etal~\cite{Hata19a}.

\finding{4}{The trend of analyzing multiple software systems of a programming language, or of several languages, shows the increasing use of polyglot environments in software projects.}

\subsection{RQ$_2$: Which \qas are used to assess code comments?}
\seclabel{resultsRq2}

\begin{figure*}[ht]
  \centering
  \includegraphics[width=0.75\textwidth]{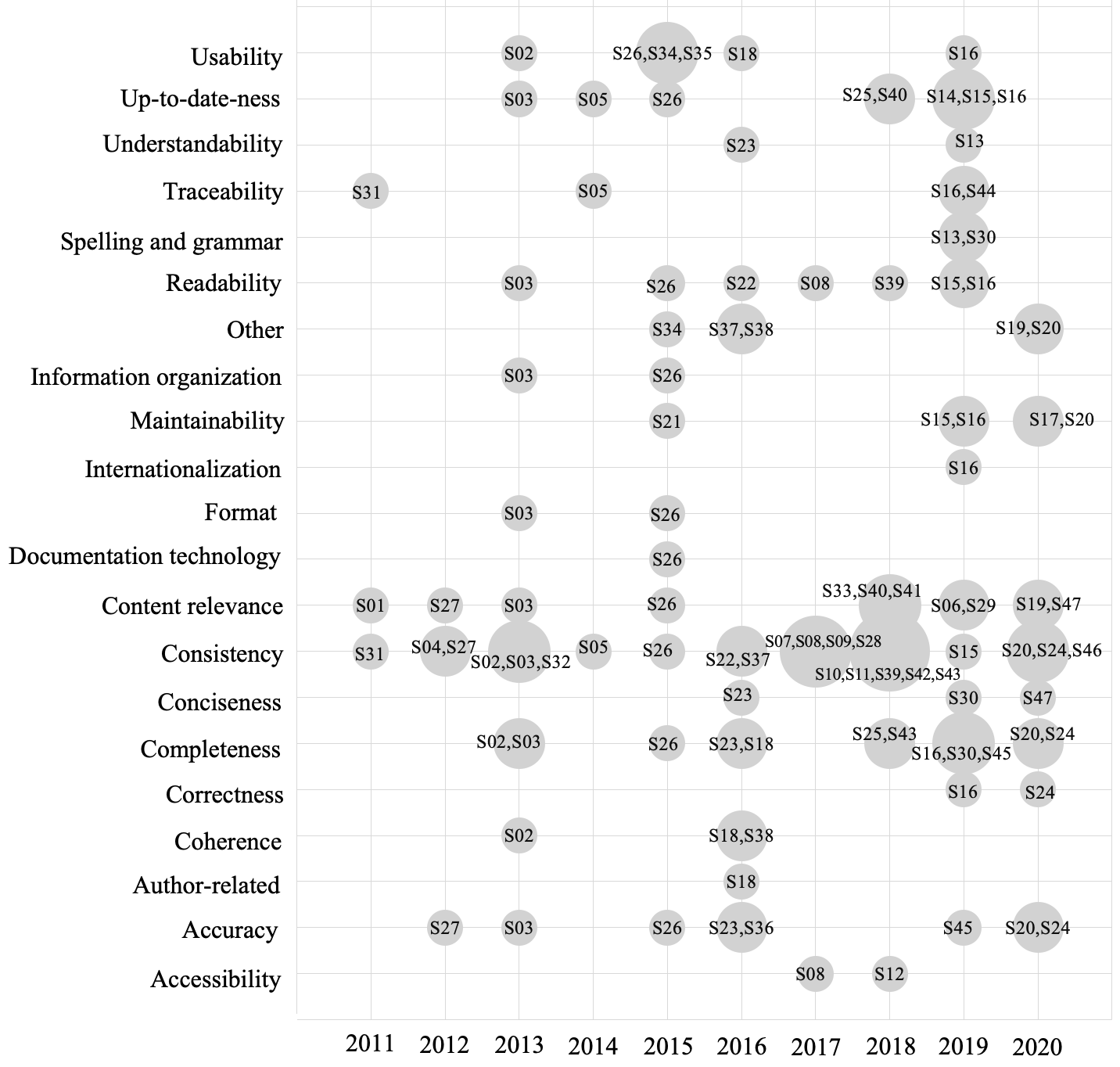}
  \caption{Frequency of various comment quality \qas over year}
  \figlabel{plot-qa-per-year}
\end{figure*}

To characterize the attention that the relevant studies reserve to each \qa over the past decade,
\figref{plot-qa-per-year} shows all the \qas on the y-axis and the corresponding years on the x-axis. Each bubble in the plot indicates both the number of papers by the size of the bubble and IDs of the studies. 
Comparing the y-axis with the \qas in \tabref{paper-fields-extraction-rq} demonstrates that our analysis finds new \qas with respect to the previous work of Zhi \etal
The \asText\additionalQA additional 
\qas are: 
\emph{usefulness}, \emph{use of examples}, \emph{usability}, 
\emph{references}, \emph{preciseness}, \emph{natural 
language quality}, \emph{maintainability}, \emph{visual models}, 
\emph{internationalization}, \emph{documentation technology}, \emph{content 
relevance}, \emph{conciseness}, \emph{coherence}, and \emph{availability}.
However, not all \qas reported by Zhi \etal for software documentation quality (highlighted in bold in \tabref{paper-fields-extraction-rq}) are used in comment quality assessment.
In particular,  we find no mention of \emph{trustworthiness}, and \emph{similarity} \qas even though previous works have highlighted the importance of both \qas to have high-quality documentation~\cite{Visc04a,Ambl07a,Daut11a}.
Also, Maalej \etal showed in their study that developers trust code comments more than other kinds of software documentation~\cite{Maal14a}, indicating the need to develop approaches to assess the trustworthiness of comments.

\finding{5}{Compared to the previous work by Zhi \etal, we find 10 additional \qas researchers use to assess code comment quality.}

Although several \qas received attention in 2013, the detailed analysis shows that there were mainly two studies (S02, S03) covering several \qas. 
There is only one study published in 2014 (S05), while 2015 sees the first studies focusing on assessing comment quality.  
One in particular, S26, attempts to cover multiple \qas.  
The plot also shows which \qas receive the most attention. A few \qas such as 
\textit{completeness}, \textit{accuracy}, \textit{content relevance}, 
\textit{readability} are often investigated. The \qa \textit{consistency} is by far the one that receives constant and consistent attention across the years, with several in 2017 (S07, S08, S09, S29) and 2018 
(S10, S11, S39, S42, S43).
Indeed, the problem  of inconsistency has been studied from multiple points of view, such as 
inconsistency between code and comments that may emerge after code 
refactoring (S07), or the inconsistencies revealed by so-called 
\textit{linguistic antipatterns} (S11, S37).
Unsurprisingly,  the plot shows that \textit{up-to-dateness} increasingly has received attention in the last three years of the decade, given that comments that are not updated together with code are also a cause of inconsistency (S15, S16).

A few attributes are rarely investigated, for instance the \qas investigated only by at most two studies over the past decade are 
\textit{format}, \textit{understandability}, \textit{spelling \& grammar}, \textit{organization}, \textit{internationalization}, \textit{documentation technology}, \textit{coherence}, \textit{conciseness},
\textit{author related} and \textit{accessibility}. More research would be needed to assess whether such attributes are intrinsically less important than others for comments according to practitioners.

\finding{6}{
  While \qas such as \emph{consistency} and \emph{completeness} are frequently used to assess comment quality, others are rarely investigated, such as \emph{conciseness} and \emph{coherence}.
}

Another aspect to analyze is whether researchers perceive the \qas as being the same or not. For example, do all studies mean the same by consistency, conciseness, accuracy of comments? 
We therefore collect the definition of each \qa considered in the study. 
We find that for various \qas researchers refer to the same \qa but using different terminology. 
We map such cases to the \emph{Synonyms} column presented in \tabref{paper-fields-extraction-rq}.
From this analysis we find that not all studies precisely define the \qas, or they refer to their existing definitions while evaluating comments using them.
For instance, the studies (S01, S04, S13, S17, S20, S29, S41) do not mention the specific \qas or their definition.
We put such studies, classifying comment content with the aim to improve comment quality, under \emph{content relevance}.
On the other hand, in some studies researchers mention the \qas but not their definition.
For instance, S26 refers to various existing studies for the \qa definitions but which \qa definition is extracted from which study is not very clear.
Lack of precise definitions of \qas or having different definitions for the same \qas can create confusion among developers and researchers while assessing comment quality.
Future work needs to pay attention to either refer to the existing standard definition of a \qa or define it clearly in the study to ensure the consistency and awareness across developer and scientific communities.
In this study, we focus on identifying the mention of \qas and their definition if given, and not on comparing and standardizing their definition. Such work would require not only the existing definitions available in the literature for \qas but also collecting how researchers use them in practice, and what developers perceive from each \qa for source code comments, which is out of scope for this work.
However, we provide the list of \qas researchers use for comment quality assessment to facilitate future work in mapping their definition and standardizing them for code comments.

Although each \qa has its own importance and role in comment quality, they are not measured in a mutually exclusive way. 
We find cases where a specific \qa is measured by measuring another \qa.
For example, \emph{accuracy} is measured by measuring the \emph{correctness} and \emph{completeness} of comment, such as ``\emph{the documentation is incorrect or incomplete and therefore no longer accurate documentation of an API.}'' (S24)
Similarly,  \emph{up-to-dateness} is measured through \emph{consistency} of comments (S40) or \emph{consistency} is evaluated and improved using \emph{traceability} (S31).
This indicates the dependency of various \qas on each other, and improving one aspect of comments can automatically improve other related aspects.
However, which techniques are used to measure which \qas is not yet known.

\finding{7}{
  Many studies miss a clear definition of the \qas they use in their studies. This poses various challenges for developers and researchers, \eg understanding what a specific \qa means, mapping a \qa to other similar \qas, and adapting the approaches to assess the \qa to a certain programming environment.
}


{\subsection{RQ$_3$: \rqIII}}
\seclabel{resultsRq3}

\begin{figure*}[tbh]
  \centering
  \includegraphics[width=0.85\textwidth]{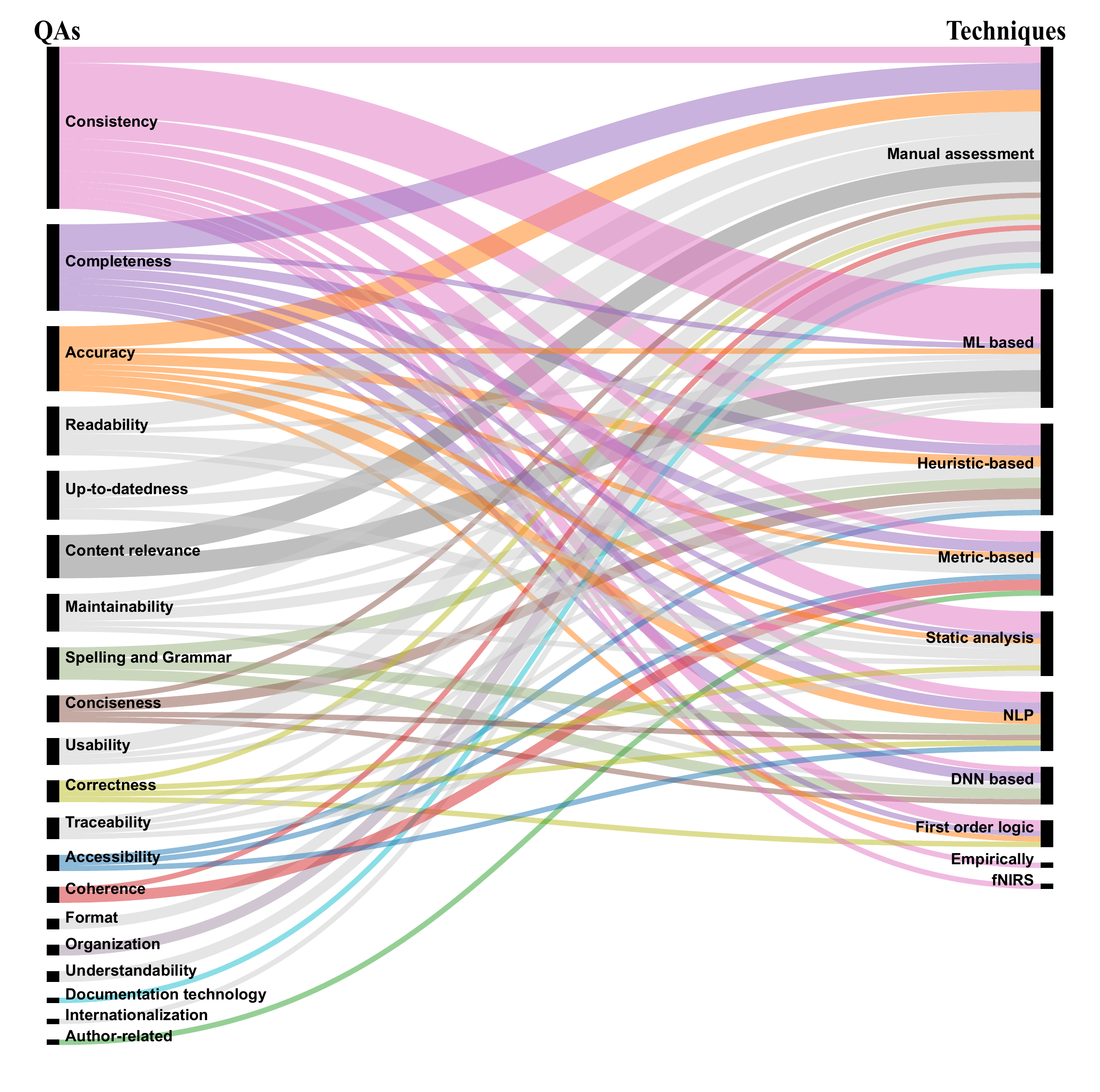}
  \caption{Types of techniques used to analyze various \qas}
  \figlabel{plot-qas-technique-types}
\end{figure*}

With respect to each \qa, we first identify which techniques have been used to measure them. We use the dimension \emph{Technique type}
to capture the type of techniques.
\figref{plot-qas-technique-types} shows that the majority of the \qas are measured by asking developers to manually assess it (\emph{manual assessment}).
For instance, \qas such as \emph{coherence}, \emph{format}, \emph{organization}, \emph{understandability}, and \emph{usability} are often assessed manually.
This indicates the need and opportunities to automate the measurement of such \qas.

A significant number of studies experimented with various automated approaches based on machine or deep learning approaches, but they focus on specific \qas and miss other \qas such as \emph{natural language quality}, \emph{conciseness}, \emph{correctness}, \emph{traceability}, \emph{coherence} \etc
Similarly, another significant portion of studies uses heuristic-based approaches to measure various \qas. 
The limitation of such heuristic-based approaches is their applicability to other software systems and programming languages. 
More studies are required to verify the generalizability of such approaches.

\finding{8}{\emph{Manual assessment} is still the most frequently-used technique to measure various \qas. 
Machine learning based techniques are the preferred automated approach to asses \qas, but the majority of them focus on specific \qas, such as \emph{consistency}, \emph{content relevance}, and \emph{up-to-dateness}, while ignoring other \qas.}

We find that the majority of the machine learning-based approaches are supervised ML approaches. 
These approaches require labeling the data and are therefore expensive in terms of time and effort.
To avoid the longer training time and memory consumption of ML strategies, Kallis \etal used \emph{fastText} to classify the issues reports on GitHub \cite{Kall21a}.
The \emph{fastText} tool uses linear models and has achieved comparable results in classification to various deep-learning based approaches.
A recent study by Minaee \etal shows that deep learning-based approaches surpassed common machine learning-based models in various text analysis areas, such as news categorization and sentiment analysis~\cite{Mina21a}. 
We also find some studies that use deep learning-based techniques partly (S06, S13, S20) along with machine learning techniques for a few \qas, such as assessing \emph{conciseness}, \emph{spelling and grammar}, and \emph{completeness}.
However, there are still many \qas that are assessed manually and require considerable effort to support developers in automatically assessing comment quality.

\finding{9}{In the case of automated approaches to assess various \qas of comments, we observe that deep-learning based approaches are not yet explored even though various studies showed that they surpassed ML-based approaches in text analysis areas.}

We see that machine learning-based approaches are used more often than deep-learning approaches, but whether it is due to their high accuracy, easy interpretation, or need for a small dataset is unclear and requires further investigation.

\begin{table*}
  \centering
  \footnotesize
  \caption{Metrics and tools used for various quality attributes. \\
  Note: the description of each metric is given in \tabref{Metrics-description}}
  \label{tab:quality-attributes-metrics-tools}
  \begin{tabular}
  {p{0.15\linewidth}p{0.25\linewidth}p{0.20\linewidth}}
  \noalign{\smallskip}\hline
  \textbf{\qas} & \textbf{Metrics} &  \textbf{Tools}  \\
  \noalign{\smallskip}\hline
        Accessibility & S08: Accessibility\_1, Accessibility\_2 & S12: Text2KnowledgeGraph \\ \hline
        Readability & S08: Readability\_1 & \\
        &  S22: Readability\_1 & \\ 
        &  S39: Readability\_1 & \\ \hline
        Spelling and Grammar & S13: SpellGrammar\_1 & ~ \\ \hline
        Correctness & ~ & S24: Drone \\ \hline
        Completeness & S18: Completeness\_1,  Author\_1 & S24: Drone \\ 
        & S02: Completeness\_2 &  S45: DAAMT \\
        & S43: Completeness\_3 &   \\ \hline
        Consistency & S08: Consistency\_1 & S07: Fraco \\ 
        & S22: Consistency\_1 & S06: RecoDoc, AdDoc \\ 
        & S39: Consistency\_1 &  S24: Drone \\ 
        & S46: Consistency\_2 & S31: Coconut \\
        & S28: LicenseConsistency\_1 & S09: Zhou et. al \\ 
        & & S37: LAPD \\
        & & S42: PyID \\ \hline
        Traceability & ~ & S06: RecoDoc \\ 
        & & S31: Coconut \\ \hline
        Up-to-datedness & ~ & S06: RecoDoc, AdDoc \\ 
        & & S14: Trigit \\ \hline
        Accuracy & S36:  Accuracy\_1 & S24: Drone \\ 
        & & S45: DAAMT \\ \hline
        Coherence & S02. Coherence\_1, Coherence\_2 \\ 
        & S18: Coherence\_3 & \\ 
        & S38: Coherence\_4 & ~ \\ \hline
        Maintainability & S21: Coherence\_4  & S17: Pham et. al. \\ \hline
        Understandability & S13: Understandability\_1 & ~ \\ \hline
        Usability & S35: Usability\_1 & S34: Krec \\
  \noalign{\smallskip}\hline
  \end{tabular}
\end{table*}

\begin{table*}[!ht]
  \centering
  \footnotesize
  \caption{Description of each metric listed in \tabref{quality-attributes-metrics-tools}}
  \tablabel{Metrics-description}
  \begin{tabular}
    {p{0.15\linewidth}p{0.80\linewidth}}
  \hline
  \textbf{Metrics} & \textbf{Description} \\ \hline
      Accessibility\_1 & MIDQ: (Documentable items of a method + readability of comments)/2. [S08] \\ \hline
      Accessibility\_2 & AEDQ: Identify all Stack overflow discussions that have "how to"  words and the class name in the title. [S08]\\ \hline
      Accuracy\_1 & Short Text Semantic Similarity (STSS): the intersection of keywords between summaries and source code, STASIS (word semantic similarity, sentence semantic similarity, and word order similarity), LSS (Lightweight Semantic Similarity). [S36] \\ \hline
      Author\_1 & A class comment should contain authorship. Check the presence and absence of the @author tag with the following name. [S18]\\ \hline
      Coherence\_1 & The similarity between words from method comments and method names where similarity is computed using Levenshtein distance.
      The value should be between 0 and 0.5 to have a coherent comment. [S02] \\ \hline
      Coherence\_2 & The length of comments should be between 2 words to 30 words. [S02] \\ \hline
      Coherence\_3 & Percentage of the number of class or method's words contained in the class or method comments divided by the total class or method's words. The value should be above or equal to 0.5. [S18]\\ \hline
      Coherence\_4 & There is coherence between the comment and the implementation of a method when they have a high lexical similarity, where lexical similarity is computed using cosine similarity. [S38] \\ \hline
      Completeness\_1 & A class comment should contain a description and authorship. A method should contain comments if it is complex (more than three method invocation) and have 30 LOC. [S18] \\ \hline
      Completeness\_2 & How many of the public classes, types, and methods have a comment preceding them. [S02] \\ \hline
      Completeness\_3 & Exceptions that are present in App Programs, Crashes, and API source code but not in API reference documentation. [S43] \\ \hline
      Consistency\_1 & The overlap between the terms used in a method comment and the terms used in the method body. They correlate a higher value of CIC with a higher readability level of that code. [S08, S22, S39] \\ \hline
      Consistency\_2 & The Kullback-Leibler divergence is a measure that finds the difference between two probability distributions. [S46]\\ \hline
      LicenseConsistency\_1 & Two similar source code files have different licenses. Find the number of files in a group, number of different licenses in the group, number of files with an unknown license in the group, number of files without any license in the group, and number of licenses in the GPL family. [S28]\\ \hline
      Readability\_1 & Flesch reading-ease test. [S08, S22, S39] \\ \hline
      SpellGrammar\_1 & The sentence has no subject or predicate, or has incomplete punctuations (e.g., the right parenthesis is missing). [S13] \\ \hline
      Understandability\_1 & Remove a sentence if it is incomplete, contains code elements, is a question, or it mentions the concept in its subordinate clauses. [S13] \\ \hline
      Usability\_1 & ADI: number of words in the method comments. The threshold is decided based on the simple average of the ADI for all method declaration. [S35]\\ \hline
  \end{tabular}
\end{table*}

In addition to identifying general techniques, we collect which metrics and tools have been used to measure various \qas. \tabref{quality-attributes-metrics-tools} shows various \qas in the column \emph{\qas}, and metrics and tools used for each \qa in the column \emph{Metrics}, and \emph{Tools} respectively. The description of the collected metrics is presented in \tabref{Metrics-description}. We can see that out of \numQA, only 10 \qas have metrics defined for them.

A software metric is a function that takes some software data as input and provides a numerical value as an output.
The output provides the degree to which the software possesses a certain attribute affecting its quality~\cite{IEEE93a}.
To limit the incorrect interpretation of the metric, threshold values are defined.
However, the threshold value may change according to the type of comments analyzed, and the interpretation of the metric output may vary in turn. We report threshold values, if present, for the collected metrics.

For \emph{readability} \qa, researchers were often found to be using the same metric (S08, S22, S39).
As developers spend significant amount of time reading code, including comments, having readable comment can help them in understanding code easier. Yet readability remains a subjective concept. 
Several studies, such as S08, S22, S39 identified various syntactic and textual features for source code and comments. 
However, in context of code comments, they focus on the Flesch-Kincaid index method, which is typically used to assess readability of natural language text. Since comments often consist of a mix of source code and natural language text, such methods can have disadvantages. For example, developers can refer to the same code concept differently in comments, and they can structure their information differently.
Thus, formulating metrics that consider the special context of code comments can improve the the assessment of readability of comments.

Another popular metric is \emph{Consistency\_1} used for assessing consistency between comments and code (S08, S22, S39). This metric measures the overlap between the terms of method comments and method body. These studies assume that the higher the overlap, better the readability of that code. Similarly, metrics (\emph{coherence\_1} , \emph{coherence\_3}, \emph{coherence\_4}) used for measuring the \emph{coherence} \qa suggest higher overlap between comments and code. However, having too many overlapping words can defy the purpose of comments and can lead to redundant comments. Using such metrics, a comment containing only rationale information about a method or class might be qualified as an incoherent or inconsistent comment whereas such comments can be very helpful in providing additional important information.
Although metrics can help developers easily estimate the quality of comments, their sensitivity towards various \qas can degrade comment quality overall.
More research is required to know the implication of given metrics on various \qas or combinations of \qas.

\finding{10}{Nearly 25\% of the studies use metric-based methods to measure comment quality. However, the metrics are defined or used for only 10 \qas out of \asText\numQA \qas.}

{\subsection{RQ$_4$: \rqIV}}
\seclabel{resultsRq4}
\begin{figure*}[tbh]
  \centering
  \includegraphics[width=\textwidth]{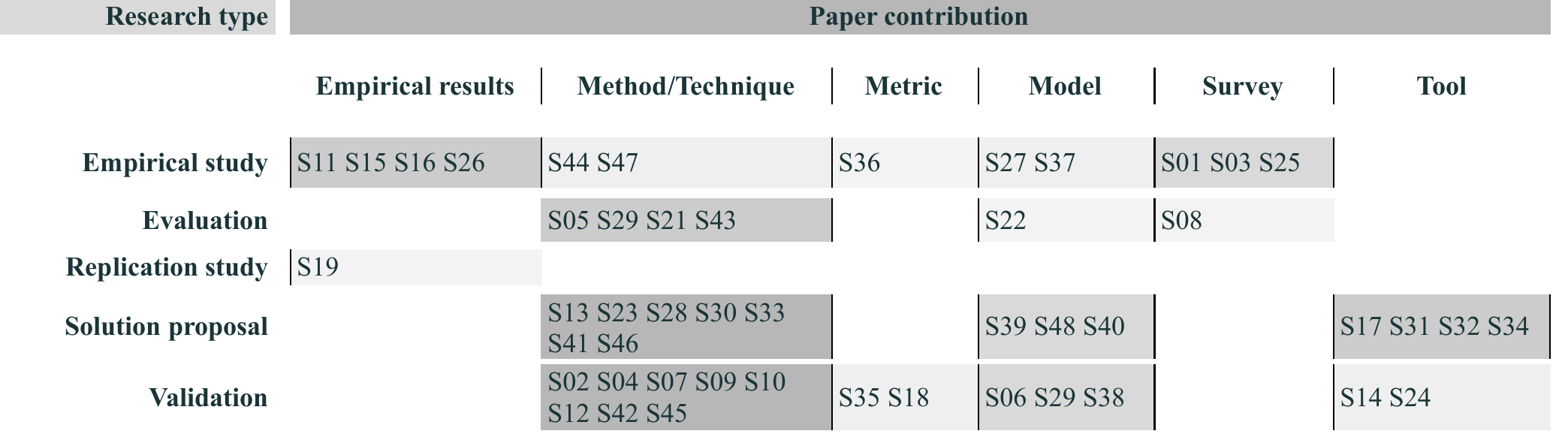}
  \caption{Types of contribution for each research type}
  \figlabel{plot-papers-research-approaches}
\end{figure*}

\begin{figure*}[tbh]
  \centering
  \includegraphics[width=\textwidth]{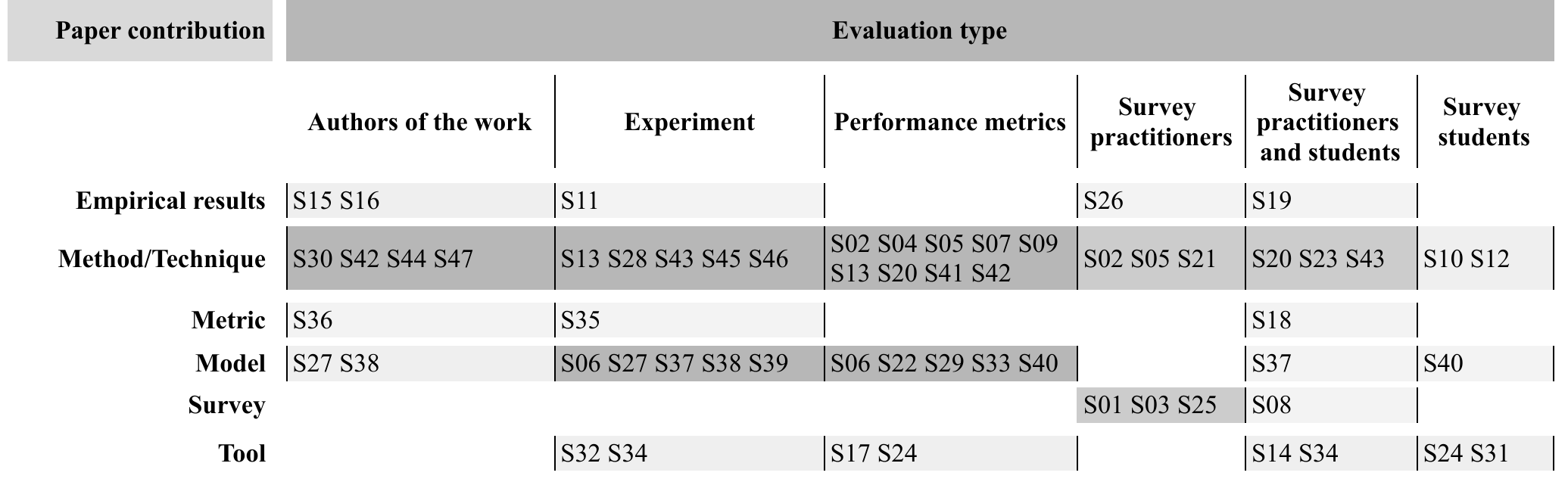}
  \caption{Types of evaluation for each paper contribution type}
  \figlabel{plot-papers-approaches-evaluations}
\end{figure*}

\paragraph{Research types}
 As a typical development cycle can contain various research tasks, such as investigation of a problem, or validation of a solution, we collect which types of research are performed for the comment quality assessment domain, and what kinds of solutions researchers often contribute.
 We categorize the papers according to the \emph{research type} dimension and show its results in \figref{plot-papers-research-approaches}.
 The results show that the studies often conduct validation research (investigating the properties of a solution) followed by the solution proposal (offering a proof-of-concept method or technique).
 However, very few studies focus on evaluation research (investigating the problem or a technique implementation in practice). 
 We find only one study performing a replication study (S19).
 Given the importance of research replicability in any field, future work needs to focus more on evaluating the proposed solution and testing their replicability in this domain.

\paragraph{Paper contribution types} 
By categorizing the papers according to the \emph{paper contribution} definition, \figref{plot-papers-research-approaches} and \figref{plot-papers-approaches-evaluations} 
show that over 44\% of papers propose an approach (method/technique) to assess code comments.
A large part (\asText{\FPuse{round(3/4*100, 0)}\%}) of them are heuristics-based approaches, \eg Zhou \etal and Wang \etal present such NLP based heuristics (S9, S13).
A few approaches rely on manual assessments. 
As an example, consider how taxonomies assessing comment quality have emerged~\cite{Wen19a, Agha19a}.
Models are the second contribution by frequency, which makes sense considering the increasing trend of leveraging machine learning during the considered decade: \asText{\FPuse{round(3/5*100, 0)}\%} of the relevant papers proposing models are based on such approaches. The label \emph{Empirical results} comprises studies which mainly offer insights through authors' observations (\eg S11, S15, S16, S19, S26).
Finally, given the important role that metrics have in software engineering~\cite{Fent14a, Mene12a}, it is valuable to look into metrics that are proposed or used to assess code comment quality as well. 
For example, three studies (S18, S35, and S36) contribute metrics for \emph{completeness}, \emph{accuracy}, or \emph{coherence} whereas other studies use existing established metrics, \eg S08, S22, or S39 compute the \emph{readability} of comments using the metric named the Flesch-Kincaid index.

\paragraph{Tool availability}
Previous work indicates the developers' effort in seeking tools to assess documentation quality, and highlights the lack of such tools ~\cite{Agha19a}.
In our study, we find that \asText{\FPuse{round(numPaperProposeTool/\numPapersRelevant*100, 0)}\%} of the studies propose tools to assess specific \qas, mainly for detecting inconsistencies between code and comments. 
Of these studies proposing tools, 
\asText{\FPuse{round(numPaperAvailableTool/numPaperProposeTool*100, 0)}\%}
provide a link to them. The lack of a direct link in the remaining
\asText{\FPuse{round(100-numPaperAvailableTool/numPaperProposeTool*100, 
		0)}\% } can hinder the reproducibility of such studies.

\paragraph{Dataset availability}
In terms of dataset availability, 
\asText{\FPuse{round(numPaperAvailableDataset/numPapersRelevant*100, 0)}\% } of the studies provide a link to a replication package. 
Of the remaining papers, some provide a link to the case studies they analyze (typically open-source projects)~\cite{Haou11a}, build on previously existing datasets~\cite{Scal18a}, or mention the reasons why they could not provide a dataset. 
For instance, Garousi \etal indicated the company policy as a reason to not to share the analyzed documentation in their case study~\cite{Garo15a}.

\finding{11}{
  Nearly 50\% of the studies still are lacking on the \emph{replicability} dimension, with their respective dataset or tool often not publicly accessible.
}


{\subsection{RQ$_5$: \rqV}}
\seclabel{resultsRq5}

\figref{plot-papers-approaches-evaluations} shows how authors evaluate their contributions.
We see that code comment assessment studies generally lack a systematic evaluation, surveying only students, or conducting case studies on specific projects only.
Most of the time, an experiment is conducted without assessing the results through any kind of external expertise judgment.
Hence, only 30\%
 of the relevant studies survey practitioners to evaluate their approach. 
This tendency leads to several disadvantages.
First, it is difficult to assess the extent to which a certain approach may overfit specific case studies while overlooking others.
Second, approaches may be unaware of the real needs and interests of project developers. 
Finally, the approaches may tend to focus too little on real-world software projects (such as large software products evolving at a fast pace in industrial environments).
Similarly, when a new \emph{method} or \emph{technique} or comment classification \emph{model} is proposed, it is often assessed based on conventional performance metrics, such as Precision, Recall, or F1 (S02, S04, S07, S29, S41 \etc) and rarely are the results verified in an industry setting or with practitioners. 

\finding{12}{Many code comment assessment studies still lack systematic 
industrial evaluations for their proposed approaches, such as evaluating the  
\emph{metric}, \emph{model}, or \emph{method/technique} with 
practitioners.}


 \section{Discussion}
 \seclabel{discussion}

Below we detail our observations about state of the art in comment quality analysis together with implications and suggestions for future research.

\paragraph{Comment Types}
The analysis of the comment quality assessment studies in the last decade shows that the trend of analyzing comments from multiple languages and systems is increasing compared to the previous decade where a majority of the studies focus on one system \cite{Zhi15a}.
It reflects the increasing use of polyglot environments in software
development \cite{Toma14a}.
Additionally, while in the past researchers focused on the quality of code comments in general terms, there is a new trend of studies that narrow their research investigation to particular comment types (methods, TODOs, deprecation, inline comments), indicating the increasing interest of researchers in supporting developers in providing a particular type of information for program comprehension and maintenance tasks.

\paragraph{{Emerging \qas}} %
Our analysis of the last decade of studies on code comment assessment shows that new \qas (\emph{coherence}, \emph{conciseness}, \emph{maintainability}, \emph{understandability} \etc), which were not identified in previous work \cite{Zhi15a}, are now being investigated and explored by researchers. 
This change can be explained by the fact that while in the past researchers focused on the quality of code comments in general terms, in the last decade there has been a new trend of studies that narrow their research investigation to specific comment types (methods, TODOs, deprecation, inline comments) and related \qas.
\paragraph{{Mapping \qas}} %
As a consequence of this shift of focus towards specific comment types, the same \qas used in prior studies can assume different definition nuances, depending on the kind of comments considered. 
For instance, let us consider how the \qa \emph{up-to-dateness}, referred to in studies on code-comment \emph{inconsistency}, assumes a different interpretation in the context of TODO comments. 
A TODO comment that \emph{becomes outdated} describes a feature that is not being implemented, which means that such a comment should be addressed within some deadline, and then removed from the code base (S14) when either the respective code is written and potentially documented with a different comment, or the feature is abandoned altogether.
At the same time, more research nowadays is conducted to understand the relations between different \qas.

\paragraph{Mapping taxonomies} 
In recent years, several taxonomies concerning code comments have been proposed, however, all of them are characterized by a rather different focus, such as the scope of the comments (S02), the information embedded in the comment (S29, S41), the issues related to specific comment types (S06, S33, S40 ), as well as the programming language they belong to. 
This suggests the need for a comprehensive code comment taxonomy or model that maps all these aspects and definitions in a more coherent manner to have a better overview of developer commenting practices across languages.
Rani \etal adapted the code comment taxonomies of Java and Python (S29, S41) for class comments of Java and Python~\cite{Rani21d}.
They mapped the taxonomies to Smalltalk class comments and found that developers write similar kinds of information in class comments across languages. 
Such a mapping can encourage building language-independent approaches for other aspects of comment quality evaluation.


\section{Implication for Future studies}
\seclabel{implication-future-work}
Besides the aspects discussed above, future studies on code comment assessment should be devoted to filling the gaps of the last decade of research as well as coping with the needs of developers interested in leveraging comment assessment tools in different program languages.

\paragraph{Investigating specific comment types (RQ1)} 
Several works showed the importance of different types of comments to achieve specific development tasks and understanding about code.  
Although, the trend of analyzing specific comment types has increased over the last decade, there are still comment types (\eg class and package comments) that need more attention.

\paragraph{Generalizing across languages (RQ1)} 
Given the preponderance of studies focusing on the Java language, and considering that statistics from various developer boards (StackOverflow, GitHub) suggest that there are other popular languages as well (\eg Python and JavaScript), more studies on analyzing various types of comments in these languages are needed.
Interesting questions in this direction could concern the comparison of practices (\eg given Python is often considered to be ``self-explainable'', do developers write fewer comments in Python?) and tools used to write code comments in different languages (\eg popularity of Javadoc v.s. Pydoc).
Similarly, whether various programming language paradigms, such as functional versus object-oriented languages, or statically-typed versus dynamic-typed languages, play a role in the way developers embed information in comments, or the way they treat comments, needs further work in this direction.

\paragraph{Identifying \qas (RQ2)} 
Our results show various \qas, \eg \emph{consistency}, \emph{completeness}, and \emph{accuracy} that are frequently considered in assessing comment quality.
Additionally, various metrics, tools, and techniques that are proposed to assess them automatically.
Indeed, some \qas are largely overlooked in the literature, \eg there is not enough research on approaches and automated tools that ensure that comments are \textit{accessible}, \textit{trustworthy}, and \textit{understandable}, despite numerous studies suggesting that having good code comments brings several benefits.

\paragraph{Standardizing \qas (RQ2)}
We identify various \qas that researchers consider assessing comment quality. Not all of these \qas are unique \ie they have conceptual overlap (based on their definitions in \tabref{paper-fields-extraction-rq} and measurement techniques in \tabref{quality-attributes-metrics-tools}).
For example, the definition of \emph{up-to-datedness} and \emph{consistency} mention of keeping comments updated. Similarly, the definition of \emph{coherence} and \emph{similarity} focus on the relatedness between code and comments.
In this study, we mainly focus on identifying various \qas from the literature and on extracting metrics, tools, and techniques to measure them.
Standardizing their definition can be an essential next step in the direction of comment quality assessment research. Since not every study provides the definition of mentioned \qas, such a work will require surveying the authors to understand how they perceive various \qas and where they refer to for \qas definitions.

\paragraph{Comment smells (RQ2)}
Although there is no standard definition of good or bad comments, many studies indicate bloated comments (or non-informative comments), redundant comments (contain same information as in the code), or inconsistent comments (\eg contain conflicting information compared to the code) as code or comment smells.
Arnaoudva \etal identified various LAs that developers perceive as poor practices and should be avoided~\cite{Arna16a}. 
Still, what information is vital in comments is a subjective concept and can sometimes be contradictory.
For instance, Oracle's coding style guideline suggests including author information in class comments, whereas the Apache style guideline suggests removing it as it can be inferred from the version control system~\cite{Orac20a}.
We find that researchers use the \emph{completeness} \qa to identify informative comments. 
They define various metrics to assess the completeness of comments, as shown in \tabref{Metrics-description}.
These metrics check the presence of specific information, such as summary, author, or exception information in class or method comments
Future work can investigate the definition of good and bad comments by surveying various sources, such as documentation guidelines, researchers, and developers, and comparing the sources across to improve the understanding of high-quality comments.  
Such work can inspire the development of more metrics and tools to ensure the adherence of comments to the standards. 

\paragraph{Automated tools and techniques (RQ3)} 
Finally, concerning techniques to assess comment quality, we observed that those based on AI, such as NLP and ML, were increasingly used in the past decade.
On the other hand, deep learning techniques do not yet seem to have gained a foothold within the community for assessing comment quality.
 Since code comment generation is becoming more and more popular also due to such advanced techniques emerging, we envision that future work may study techniques and metrics to assess the quality of automatically generated code comments.

\paragraph{Research evaluation (RQ4 and RQ5)}
Scientific methods play a crucial role in the growth of engineering knowledge~\cite{Vinc90a}. 
Several studies have indicated the weak validation in software engineering~\cite{Zelk97a}. 
We also find that several studies propose solutions but do not evaluate their solution. 
Also, various approaches were validated only by the authors of the work or by surveying students. 
However, we need to do all steps as engineering science researchers do, empirically investigating the problems, proposing solutions, and validating those solutions.

In contrast to seven research types listed in \autoref{tab:paper-research-contribution-type}, we observe only limited types of research studies.
For example, we do not find any philosophical, opinion, or experience papers for the comment quality assessment domain even though this domain is more than a decade old now.
Philosophical papers sketch a new perspective of looking at things, conceptual frameworks, metrics \etc
Opinion papers present good or bad opinions of authors about something, such as different approaches to assess quality, using particular frameworks \etc 
Similarly, experience papers often present insights about lessons learned or anecdotes by authors in using tools or techniques in practice. Such papers help tool designers better shape their future tools.

\section{Threats to validity}
\seclabel{Threats-to-validity}
We now outline potential threats to the validity of our study.
\emph{Threats to construct validity}
mainly concern the measurements used in the evaluation process.
In this case, threats can be mainly due to (i) the imprecision in the automated selection of relevant papers (\ie the three-step search on the conference proceedings based on regular expressions), and to (ii) the subjectivity and error-proneness of the subsequent manual classification and categorization of relevant papers. 

We mitigated the first threat by manually classifying a sample of relevant papers from a set of conference proceedings and compared this classification with the one recommended by the automated approach based on regular expressions. 
This allowed us to incrementally improve the initial set of regular expressions.
To avoid any bias in the selection of the papers, we selected regular expression in a deterministic way (as detailed in the \secref{study-design}): We first examined the definition of \emph{documentation} and \emph{comment} in \emph{IEEE Standard Glossary of Software Engineering Terminology} (IEEE Standard 610.12-1990) and identified the first set of keywords \emph{comment},
\emph{documentation}, and \emph{specification};  we further added comment-related 
keywords that are frequently mentioned in the context of code comments.   
Moreover, we formulated a set of keywords to discard irrelevant studies that presented similar keywords (e.g., code review comments).  
To verify the correctness of the final set of keywords, we manually scanned the full venue proceedings metadata to make sure the set of keywords did not prune relevant papers. This iterative approach allowed us to verify that our keyword-based filtering approach does not lead to false negatives for
the selected venues.

We mitigated the second threat by applying multi-stage manual classification of conference proceedings, involving multiple evaluators and reviewers, as detailed in \secref{study-design}.

\emph{Threats to internal validity} concern confounding factors that could influence our results and findings. 
A possible source of bias might be related to the way we selected and analyzed the conference proceedings. 
To deal with potential threats regarding the actual regular expressions considered for the selection of relevant studies, we created regular expressions that tend to be very inclusive, \ie that
select papers that are marginally related to the topic of interest, and we take a final decision only after a manual assessment.

\emph{Threats to external validity} concern the generalization and completeness of results and findings. 
Although the number of analyzed papers is large, since it involves studies spanning the last ten years of research, there is still the possibility that we missed some relevant studies.
We mitigate this threat by applying various selection criteria to select relevant conference proceedings, considering the well-established venues and communities related to code comment-related studies, as detailed in \secref{study-design}.
It is important to mention that this paper intentionally limits its scope in two ways, which threatens to the completeness of the study results and findings.  
First of all, we mainly focus on research work investigating code comment quality without integrating studies from industry tracks of conference venues (as was done in previous studies thematically close to ours \cite{Ding14a,Zhi15a}). 
Second, we focus on those studies that involve manually written code comments in order to avoid auto-generated comments (already investigated in recent related work \cite{Song19a,Naza16a}). 
To further limit potential threats concerning the completeness of our study, we use the snowball approach to reach potentially relevant studies that we could
have missed with our venue selection.
However, we support the argument of Garousi {\em \etal}~\cite{Garo16a} who report that a \emph{multivocal} literature review, with further replications, is desirable to make the overall interpretation of code comment quality attributes more complete for future work.


\section{Related Work}
\seclabel{Related-work}

This section discusses the literature concerning  (i) studies motivating the importance of quality attributes for software documentation, (ii) comment quality aspects, and (iii) 
recent SLRs discussing topics closely related to our investigation.

\textbf{Important quality attributes for software documentation.} 
Various research works conducted surveys with developers to identify important quality attributes of good software documentation.
Forward and Lethbridge surveyed 48 developers, and highlighted developer concerns about outdated documentation~\cite{Forw02a}.
Chen and Huang surveyed 137 project managers and software engineers~\cite{Chen09a}.
Their study highlighted the typical quality problems developers face in maintaining software documentation: adequacy, complete, traceability, consistency, and trustworthiness.
Robillard \etal conducted personal interviews with 80 practitioners and presented the important attributes for good documentation, such as including examples and usage information, complete, organized, and better design~\cite{Robi09a}.
Similarly, Plosch \etal surveyed 88 practitioners and identified consistency, clarity,
accuracy, readability, organization, and understandability as the most important attributes~\cite{Plos14a}.
They also indicated that developers do not consider documentation standards important (\eg ISO 26514:2008, IEEE Std.1063:2001).
Sohan \etal in their survey study highlighted the importance of examples in documentation~\cite{Soha17a}.
The majority of the highlighted documentation quality attributes apply to code comments as well (as a type of software documentation).
However, which specific quality attributes (\eg outdated, complete, consistent, traceable) researchers consider important to assess code comment quality and how these quality attributes are measured is yet to study.

\textbf{Comment quality.}
Evaluating comment quality according  to various aspects has gained a lot of attention from researchers,  for instance, assessing their adequacy~\cite{Arth89a} and their content quality~\cite{Kham10a,Stei13b}, analyzing co-evolution of comments and code~\cite{Flur09a}, or detecting inconsistent comments~\cite{Rato17a,Wen19a}. Several works have proposed tools and techniques for the automatic assessment of comment quality~\cite{Kham10a,Stei13b,Yu16a}.
For instance, Khamis \etal assessed the quality of inline comments based on consistency and language quality using a heuristic-based approach~\cite{Kham10a}.
Steidl \etal evaluated documentation comment quality based on four quality attributes, such as consistency, coherence, completeness, and usefulness of comments using a machine learning-based model~\cite{Stei13b}.
Zhou \etal proposed a heuristic and natural language processing-based technique to detect incomplete and incorrect comments~\cite{Zhou17a}.
These works have proposed various new quality attributes to assess comment quality, such as completeness, coherence, and language quality, that are not included in previous quality models.
However, a unifying overview of comment \qas and their assessment approaches is still missing. Our paper complements these previous works by investigating comment \qas discussed in the last decade of research.

\textbf{Previous SLRs on code comments and software documentation.}
In recent years, SLRs have been conducted to investigate agile software development aspects in open-source projects \cite{Silv17a}, the usage of ontologies in software process assessment \cite{Tarh17a}, and improvement aspects in DevOps process and practices \cite{Bads20a}.
Previous SLRs in the field investigated code comments and software documentation \cite{Ding14a,Zhi15a}, which are closely related to our work.
Specifically, Ding \etal conducted an SLR to explore the usage of knowledge-based approaches in software documentation~\cite{Ding14a}.
They identified twelve \qas.
They also highlighted the need to improve \qas, especially conciseness, credibility, and unambiguity.
Zhi \etal have explored various types of software documentation to see which \qas impact it~\cite{Zhi15a}.
Both of the studies considered the timeline until \yearBegin.
Additionally, they have not studied how the proposed comment quality
assessment approaches are computed in practice for comments.
Inspired by these related studies, we focused specifically on the code comment aspect.
Song \etal conducted a literature review on code comment generation
techniques, and indicated the need to design an objective comment
quality assessment model~\cite{Song19a}. 
Complementarily, Nazar \etal~\cite{Naza16a} presented a literature review in the field of summarizing software artifacts, which included  source code comment generation as well as bug reports, mailing lists, and developer discussion artifacts.
Our work complements these previous studies since we mainly focus on manually written comments.

\section{Conclusion}
\seclabel{conclusion}
In this work, we present the results of a systematic literature review on source code comment quality evaluation practices in the decade
\asText\yearBegin --- \asText\yearEnd.  
We study \asText\numPapersRelevant publications to understand of effort of Software Engineering researchers, in terms of what type of comments they focus their studies on, what \qas they consider relevant, what techniques they resort to in order to assess their \qas, and finally, how they evaluate their contributions.
Our findings show that most studies consider only comments in Java source files, and thus may not generalize to comments of other languages, and they focus on only a few \qas, especially on consistency between code and comments.
Some \qas, such as conciseness, coherence, organization, and usefulness, are rarely investigated.  
As coherent and concise comments play an important role in program understanding, establishing approaches to assess these attributes requires more attention from the community.
We also observe that the majority of the approaches appear to be based on heuristics rather than machine learning or other techniques and, in general, need better evaluation. 
Such approaches require validation on other languages and projects to generalize them.  
Though the trend of analyzing comments appearing in multiple projects and languages is increasing compared to the previous decade, as reported by Zhi \etal, the approaches still need more thorough validation\cite{Zhi15a}.  

 \section*{Acknowledgement}
 \seclabel{Acknowledgement}
 We gratefully acknowledge the financial support of the Swiss National Science Foundation for the project
``Agile Software Assistance'' (SNSF project No.\ 200020-181973, Feb 1,
2019 - Apr 30, 2022) and the Spanish Government through the SCUM grant
RTI2018-102043-B-I00, and the Madrid Regional through the project BLOQUES. 
We also acknowledge the Horizon 2020 (EU Commission) support for the project \textit{COSMOS} (DevOps for Complex Cyber-physical Systems), Project No. 957254-COSMOS.

\bibliographystyle{elsarticle-num}
\bibliography{scg}

\end{document}